%% file: SideChAtkStConve.tex
\documentclass[conference,a4paper]{IEEEtran}

\usepackage[dvips]{graphicx,color}
\usepackage{amsmath}
\usepackage{latexsym}
\usepackage{amssymb}
\usepackage{comment}
\usepackage{cite}
\usepackage{url}
\usepackage[T3,T1]{fontenc}

\global\long\def\L{\mathsf{L}}
\global\long\def\D{\mathsf{D}}
\global\long\def\Sgen{\mathcal{S}_\mathsf{gen}}
\global\long\def\Kgen{\mathcal{K}_\mathsf{gen}}
\global\long\def\A{\mathcal{A}}

\newcommand{\calVarX}{{\cal X}}

\newcommand{\tX}{\overline{X}}

\newcommand{\E}{\mbox{\bf E}}

\newcommand{\Dist}[1]{{#1}}

\begin{document}

\title{
Strong Converse Theorem for Source Encryption 
under Side-Channel Attacks
}
\author{%
	\IEEEauthorblockN{Yasutada Oohama and Bagus Santoso}
	\IEEEauthorblockA{University of Electro-Communications, Tokyo, Japan\\ 
	Email: \url{{oohama,santoso.bagus}@uec.ac.jp}}
%
}
\maketitle

\input{Bagus-Isit2022-abstract.tex}

\newcommand{\OmitZero}{
We pose and investigate the security problem of a source encryption 
using common key under the side channel attacks. This cryptosystem 
includes the secrecy problem for encrypted sources under the side 
channel attacks, which was previously studied by Santoso and Oohama. In 
this paper we propose a new security criterion based on the mutual 
information between the plaintext and the ciphertext. Under this 
criterion, we establish the necessary and sufficient condition for the 
secure transmission of information sources.
}

%

\newcommand{\qed}{\hfill$\square$}
\newcommand{\suchthat}{\mbox{~s.t.~}}
\newcommand{\markov}{\leftrightarrow}

\newcommand{\argmax}{\mathop{\rm argmax}\limits}
\newcommand{\argmin}{\mathop{\rm argmin}\limits}

\newcommand{\ExP}{\rm e}

\newcommand{\Cls}{class NL}
\newcommand{\vSpa}{\vspace{0.3mm}}
\newcommand{\Prmt}{\zeta}
\newcommand{\pj}{\omega_n}

\newfont{\bg}{cmr10 scaled \magstep4}
\newcommand{\bigzerol}{\smash{\hbox{\bg 0}}}
\newcommand{\bigzerou}{\smash{\lower1.7ex\hbox{\bg 0}}}
\newcommand{\nbn}{\frac{1}{n}}
\newcommand{\ra}{\rightarrow}
\newcommand{\la}{\leftarrow}
\newcommand{\ldo}{\ldots}
\newcommand{\typi}{A_{\epsilon }^{n}}
\newcommand{\bx}{\hspace*{\fill}$\Box$}
\newcommand{\pa}{\vert}
\newcommand{\ignore}[1]{}


\newtheorem{proposition}{Proposition}
\newtheorem{definition}{Definition}
\newtheorem{theorem}{Theorem}
\newtheorem{lemma}{Lemma}
\newtheorem{corollary}{Corollary}
\newtheorem{remark}{Remark}
\newtheorem{property}{Property}

\newcommand{\defeq}{:=}

\newcommand{\Qed}{\hbox{\rule[-2pt]{3pt}{6pt}}}
\newcommand{\beq}{\begin{equation}}
\newcommand{\eeq}{\end{equation}}
\newcommand{\beqa}{\begin{eqnarray}}
\newcommand{\eeqa}{\end{eqnarray}}
\newcommand{\beqno}{\begin{eqnarray*}}
\newcommand{\eeqno}{\end{eqnarray*}}
\newcommand{\ba}{\begin{array}}
\newcommand{\ea}{\end{array}}

\newcommand{\vc}[1]{\mbox{\boldmath $#1$}}
\newcommand{\lvc}[1]{\mbox{\scriptsize \boldmath $#1$}}
\newcommand{\svc}[1]{\mbox{\scriptsize\boldmath $#1$}}

\newcommand{\wh}{\widehat}
\newcommand{\wt}{\widetilde}
\newcommand{\ts}{\textstyle}
\newcommand{\ds}{\displaystyle}
\newcommand{\scs}{\scriptstyle}
\newcommand{\vep}{\varepsilon}
\newcommand{\rhp}{\rightharpoonup}
\newcommand{\cl}{\circ\!\!\!\!\!-}
\newcommand{\bcs}{\dot{\,}.\dot{\,}}
\newcommand{\eqv}{\Leftrightarrow}
\newcommand{\leqv}{\Longleftrightarrow}

\newcommand{\irr}[1]{{\color[named]{Red}#1\normalcolor}}

\newcommand{\hugel}{{\arraycolsep 0mm
                    \left\{\ba{l}{\,}\\{\,}\ea\right.\!\!}}
\newcommand{\Hugel}{{\arraycolsep 0mm
                    \left\{\ba{l}{\,}\\{\,}\\{\,}\ea\right.\!\!}}
\newcommand{\HUgel}{{\arraycolsep 0mm
                    \left\{\ba{l}{\,}\\{\,}\\{\,}\vspace{-1mm}
                    \\{\,}\ea\right.\!\!}}
\newcommand{\huger}{{\arraycolsep 0mm
                    \left.\ba{l}{\,}\\{\,}\ea\!\!\right\}}}
\newcommand{\Huger}{{\arraycolsep 0mm
                    \left.\ba{l}{\,}\\{\,}\\{\,}\ea\!\!\right\}}}
\newcommand{\HUger}{{\arraycolsep 0mm
                    \left.\ba{l}{\,}\\{\,}\\{\,}\vspace{-1mm}
                    \\{\,}\ea\!\!\right\}}}

\newcommand{\hugebl}{{\arraycolsep 0mm
                    \left[\ba{l}{\,}\\{\,}\ea\right.\!\!}}
\newcommand{\Hugebl}{{\arraycolsep 0mm
                    \left[\ba{l}{\,}\\{\,}\\{\,}\ea\right.\!\!}}
\newcommand{\HUgebl}{{\arraycolsep 0mm
                    \left[\ba{l}{\,}\\{\,}\\{\,}\vspace{-1mm}
                    \\{\,}\ea\right.\!\!}}
\newcommand{\hugebr}{{\arraycolsep 0mm
                    \left.\ba{l}{\,}\\{\,}\ea\!\!\right]}}
\newcommand{\Hugebr}{{\arraycolsep 0mm
                    \left.\ba{l}{\,}\\{\,}\\{\,}\ea\!\!\right]}}
\newcommand{\HUgebr}{{\arraycolsep 0mm
                    \left.\ba{l}{\,}\\{\,}\\{\,}\vspace{-1mm}
                    \\{\,}\ea\!\!\right]}}

\newcommand{\hugecl}{{\arraycolsep 0mm
                    \left(\ba{l}{\,}\\{\,}\ea\right.\!\!}}
\newcommand{\Hugecl}{{\arraycolsep 0mm
                    \left(\ba{l}{\,}\\{\,}\\{\,}\ea\right.\!\!}}
\newcommand{\hugecr}{{\arraycolsep 0mm
                    \left.\ba{l}{\,}\\{\,}\ea\!\!\right)}}
\newcommand{\Hugecr}{{\arraycolsep 0mm
                    \left.\ba{l}{\,}\\{\,}\\{\,}\ea\!\!\right)}}

\newcommand{\hugepl}{{\arraycolsep 0mm
                    \left|\ba{l}{\,}\\{\,}\ea\right.\!\!}}
\newcommand{\Hugepl}{{\arraycolsep 0mm
                    \left|\ba{l}{\,}\\{\,}\\{\,}\ea\right.\!\!}}
\newcommand{\hugepr}{{\arraycolsep 0mm
                    \left.\ba{l}{\,}\\{\,}\ea\!\!\right|}}
\newcommand{\Hugepr}{{\arraycolsep 0mm
                    \left.\ba{l}{\,}\\{\,}\\{\,}\ea\!\!\right|}}

\newcommand{\MEq}[1]{\stackrel{
{\rm (#1)}}{=}}

\newcommand{\MLeq}[1]{\stackrel{
{\rm (#1)}}{\leq}}

\newcommand{\ML}[1]{\stackrel{
{\rm (#1)}}{<}}

\newcommand{\MGeq}[1]{\stackrel{
{\rm (#1)}}{\geq}}

\newcommand{\MG}[1]{\stackrel{
{\rm (#1)}}{>}}

\newcommand{\MPreq}[1]{\stackrel{
{\rm (#1)}}{\preceq}}

\newcommand{\MSueq}[1]{\stackrel{
{\rm (#1)}}{\succeq}}

\newcommand{\MSubeq}[1]{\stackrel{
{\rm (#1)}}{\subseteq}}

\newcommand{\MSupeq}[1]{\stackrel{
{\rm (#1)}}{\supseteq}}

\newcommand{\MRarrow}[1]{\stackrel{
{\rm (#1)}}{\Rightarrow}}

\newcommand{\MLarrow}[1]{\stackrel{
{\rm (#1)}}{\Leftarrow}}

\newcommand{\SZZpp}{
}

\newcommand{\vcc}{{c}^n}
\newcommand{\vck}{{k}^n}
\newcommand{\vcx}{{x}^n}
\newcommand{\vcy}{{y}^n}
\newcommand{\vcz}{{z}^n}
\newcommand{\vckone}{{k}_1^n}
\newcommand{\vcktwo}{{k}_2^n}
\newcommand{\vcxone}{{x}^n}

\newcommand{\vcxtwo}{{x}_2^n}
\newcommand{\vcyone}{{y}_1^n}
\newcommand{\vcytwo}{{y}_2^n}

\newcommand{\cvcx}{\check{x}^n}
\newcommand{\cvcy}{\check{y}^n}
\newcommand{\cvcz}{\check{z}^n}
\newcommand{\cvcxone}{\check{x}^n}
\newcommand{\cvcxtwo}{\check{x}_2^n}

\newcommand{\hvcx}{\widehat{x}^n}
\newcommand{\hvcy}{\widehat{y}^n}
\newcommand{\hvcz}{\widehat{z}^n}
\newcommand{\hvckone}{\widehat{k}_1^n}
\newcommand{\hvcktwo}{\widehat{k}_2^n}

\newcommand{\hvcxone}{\widehat{x}^n}
\newcommand{\hvcxtwo}{\widehat{x}_2^n}

\newcommand{\lvcc}{{c}^n}
\newcommand{\lvck}{{k}^n}
\newcommand{\lvcx}{{x}^n}
\newcommand{\lvcy}{{y}^n}
\newcommand{\lvcz}{{z}^n}

\newcommand{\lvckone}{{k}_1^n}
\newcommand{\lvcktwo}{{k}_2^n}
\newcommand{\lvcxone}{{x}_1^n}
\newcommand{\lvcxtwo}{{x}_2^n}
\newcommand{\lvcyone}{{y}_1^n}
\newcommand{\lvcytwo}{{y}_2^n}

\newcommand{\hlvckone}{\widehat{k}_1^n}
\newcommand{\hlvcktwo}{\widehat{k}_2^n}
\newcommand{\hlvcxone}{\widehat{x}^n}
\newcommand{\hlvcxtwo}{\widehat{x}_2^n}

\newcommand{\lcvcxone}{\check{x}^n}
\newcommand{\lcvcxtwo}{\check{x}_2^n}

\newcommand{\rvcc}{{C}^n}
\newcommand{\rvck}{{K}^n}
\newcommand{\rvcx}{{X}^n}
\newcommand{\rvcy}{{Y}^n}
\newcommand{\rvcz}{{Z}^n}
\newcommand{\rvccone}{{C}_1^n}
\newcommand{\rvcctwo}{{C}_2^n}
\newcommand{\rvckone}{{K}_1^n}
\newcommand{\rvcktwo}{{K}_2^n}
\newcommand{\rvcxone}{{X}^n}

\newcommand{\rvcxtwo}{{X}_2^n}
\newcommand{\rvcyone}{{Y}_1^n}
\newcommand{\rvcytwo}{{Y}_2^n}
\newcommand{\hrvcx}{\widehat{X}^n}
\newcommand{\hrvcxone}{\widehat{X}_1^n}
\newcommand{\hrvcxtwo}{\widehat{X}_2^n}
\newcommand{\crvcx}{\check{X}^n}
\newcommand{\crvcxone}{\check{X}_1^n}
\newcommand{\crvcxtwo}{\check{X}_2^n}

  \newcommand{\cvcc}{\check{c}^m}
 \newcommand{\lcvcc}{\check{c}^m}
 \newcommand{\crvcc}{\check{C}^m}
\newcommand{\lcrvcc}{\check{C}^m}
\newcommand{\lrvcc}{{C}^n}
\newcommand{\lrvck}{{K}^n}
\newcommand{\lrvcx}{{X}^n}
\newcommand{\lrvcy}{{Y}^n}
\newcommand{\lrvcz}{{Z}^n}
\newcommand{\lrvckone}{{K}_1^n}
\newcommand{\lrvcktwo}{{K}_2^n}
\newcommand{\lrvcxone}{{X}_1^n}
\newcommand{\lrvcxtwo}{{X}_2^n}
\newcommand{\lrvcyone}{{Y}_1^n}
\newcommand{\lrvcytwo}{{Y}_2^n}
\newcommand{\lhrvcx}{\widehat{X}^n}
\newcommand{\lhrvcxone}{\widehat{X}_1^n}
\newcommand{\lhrvcxtwo}{\widehat{X}_2^n}
\newcommand{\lcrvcx}{\check{X}^n}
\newcommand{\lcrvcxone}{\check{X}_1^n}
\newcommand{\lcrvcxtwo}{\check{X}_2^n}
\newcommand{\rvcci}{{C}_i^n}
\newcommand{\rvcki}{{K}_i^n}
\newcommand{\rvcxi}{{X}_i^n}
\newcommand{\rvcyi}{{Y}_i^n}
\newcommand{\hrvcxi}{\widehat{X}_i^n}
\newcommand{\crvcxi}{\check{X}_i^n}
\newcommand{\vcki}{{k}_i^n}
\newcommand{\vcsi}{{s}_i^n}
\newcommand{\vcti}{{t}_i^n}
\newcommand{\vcvi}{{v}_i^n}
\newcommand{\vcwi}{{w}_i^n}
\newcommand{\vcxi}{{x}_i^n}
\newcommand{\vcyi}{{y}_i^n}

\newcommand{\vcs}{{s}^n}
\newcommand{\vct}{{t}^n}
\newcommand{\vcv}{{v}^n}
\newcommand{\vcw}{{w}^n}
%
%

\newcommand{\SZZ}{

\newcommand{\vcc}{{\vc c}}
\newcommand{\vck}{{\vc k}}
\newcommand{\vcx}{{\vc x}}
\newcommand{\vcy}{{\vc y}}
\newcommand{\vcz}{{\vc z}}
\newcommand{\vckone}{{\vc k}_1}
\newcommand{\vcktwo}{{\vc k}_2}
\newcommand{\vcxone}{{\vc x}_1}
\newcommand{\vcxtwo}{{\vc x}_2}
\newcommand{\vcyone}{{\vc y}_1}
\newcommand{\vcytwo}{{\vc y}_2}

\newcommand{\cvcx}{\check{\vc x}}
\newcommand{\cvcy}{\check{\vc y}}
\newcommand{\cvcz}{\check{\vc z}}
\newcommand{\cvcxone}{\check{\vc x}_1}
\newcommand{\cvcxtwo}{\check{\vc x}_2}

\newcommand{\hvcx}{\widehat{\vc x}}
\newcommand{\hvcy}{\widehat{\vc y}}
\newcommand{\hvcz}{\widehat{\vc z}}
\newcommand{\hvckone}{\widehat{\vc k}_1}
\newcommand{\hvcktwo}{\widehat{\vc k}_2}
\newcommand{\hvcxone}{\widehat{\vc x}_1}
\newcommand{\hvcxtwo}{\widehat{\vc x}_2}

\newcommand{\lvcc}{{c}}
\newcommand{\lvck}{{k}}
\newcommand{\lvcx}{{x}}
\newcommand{\lvcy}{{y}}
\newcommand{\lvcz}{{z}}

\newcommand{\lvckone}{{k}_1}
\newcommand{\lvcktwo}{{k}_2}

\newcommand{\lvcxone}{{x}}

\newcommand{\lvcxtwo}{{x}_2}
\newcommand{\lvcyone}{{y}_1}
\newcommand{\lvcytwo}{{y}_2}

\newcommand{\clvcxone}{\check{x}_1}
\newcommand{\clvcxtwo}{\check{x}_2}

\newcommand{\hlvckone}{\widehat{k}_1}
\newcommand{\hlvcktwo}{\widehat{k}_2}
\newcommand{\hlvcxone}{\widehat{x}_1}
\newcommand{\hlvcxtwo}{\widehat{x}_2}

\newcommand{\rvcc}{{\vc C}}
\newcommand{\rvck}{{\vc K}}
\newcommand{\rvcx}{{\vc X}}
\newcommand{\rvcy}{{\vc Y}}
\newcommand{\rvcz}{{\vc Z}}
\newcommand{\rvccone}{{\vc C}_1}
\newcommand{\rvcctwo}{{\vc C}_2}
\newcommand{\rvckone}{{\vc K}_1}
\newcommand{\rvcktwo}{{\vc K}_2}

\newcommand{\rvcxone}{{\vc X}}

\newcommand{\rvcxtwo}{{\vc X}_2}
\newcommand{\rvcyone}{{\vc Y}_1}
\newcommand{\rvcytwo}{{\vc Y}_2}
\newcommand{\hrvcxone}{\widehat{\vc X}_1}
\newcommand{\hrvcxtwo}{\widehat{\vc X}_2}

\newcommand{\lrvcc}{{C}}
\newcommand{\lrvck}{{K}}
\newcommand{\lrvcx}{{X}}
\newcommand{\lrvcy}{{Y}}
\newcommand{\lrvcz}{{Z}}
\newcommand{\lrvckone}{{K}_1}
\newcommand{\lrvcktwo}{{K}_2}
\newcommand{\lrvcxone}{{X}_1}
\newcommand{\lrvcxtwo}{{X}_2}
\newcommand{\lrvcyone}{{Y}_1}
\newcommand{\lrvcytwo}{{Y}_2}
\newcommand{\rvcci}{{\vc C}_i}
\newcommand{\rvcki}{{\vc K}_i}
\newcommand{\rvcxi}{{\vc X}_i}
\newcommand{\rvcyi}{{\vc Y}_i}
\newcommand{\hrvcxi}{\widehat{\vc X}_i}
\newcommand{\vcki}{{\vc k}_i}
\newcommand{\vcsi}{{\vc s}_i}
\newcommand{\vcti}{{\vc t}_i}
\newcommand{\vcvi}{{\vc v}_i}
\newcommand{\vcwi}{{\vc w}_i}
\newcommand{\vcxi}{{\vc x}_i}
\newcommand{\vcyi}{{\vc y}_i}

\newcommand{\vcs}{{\vc s}}
\newcommand{\vct}{{\vc t}}
\newcommand{\vcv}{{\vc v}}
\newcommand{\vcw}{{\vc w}}
}

\newcommand{\loF}{\underline{F}}

\newcommand{\prmtA}{\mu}
\newcommand{\prmtB}{\bar{\mu}}

\input{Bagus-Isit2022-introduction.tex}
\newcommand{\DelOneA}{
\section{Introduction \label{sec:introduction}}
As more cryptographic devices are deployed in
open physical spaces,
a new security challenge has risen
in the form of attackers which launch
\emph{side-channel attacks}, where
an attacker does not only collect the encrypted
data sent to the public communication channel,
but also collect
physical information related to the secret data
which are leaked by the devices
such as power consumption,
electromagnetic radiation, running time, etc.
Therefore, 
\emph{how to design an encryption scheme
	which is guaranteed to be secure
	even under side-channel attacks}
is a very important issue.

In this paper, we propose a general framework
for analyzing \emph{any} source encryption
with a symmetric key under \emph{any} kind of
side-channel attacks where the attacker
obtains some leaked information on the secret key.
Although  Santoso and Oohama
investigated the similar problem in
\cite{santosoOh:19},
their work is  limited only to a single \emph{specific}
encryption scheme, i.e., one-time-pad encryption,
while our framework covers
\emph{any} encryption scheme.
Then, we propose a new security criterion for secrecy
which is defined as \emph{the maximum of all conditional mutual information
	between the ciphertext and plaintext given
	the adversarial key leakage, taken over
	all probability distributions of plaintexts}.
One can easily see that our new security criterion is
more strict than the standard security criterion, i.e.,
the ordinary mutual information, as it is a usual practice
to derive an upper-bound of the security criterion
in order to guarantee secrecy of an encryption scheme.
Nevertheless, we show in this paper that we can construct
a concrete encryption scheme with reliable decoding 
and secrecy under side-channel attacks on the secret key.


The most important result obtained from
the modeling of our framework
and the new security criterion is  
the strong converse,
i.e., the necessary condition  
to have an encryption scheme with both
reliable decoding and secrecy under side-channel attacks on the secret key.
At the heart of our proof of the strong converse 
is the measurement of correlation 
between the ciphertexts and the information obtained by 
the adversary via side-channel attacks,
given distribution of the plaintexts.  
}
\section{Problem Formulation}

\subsection{Preliminaries}

In this subsection, we show the basic notations and related consensus 
used in this paper. 

\noindent{}
\textit{Random Source of Information and Key: \ }
Let $X$ be a random variable from a finite set 
$\mathcal{X}$. 
Let $\{X_t\}_{t=1}^\infty$ be a stationary discrete
memoryless source(DMS) such that for each $t=1,2,\ldots$, 
$X_{t}$ takes values in finite set $\mathcal{X}$ 
and obeys the same distribution 
as that of $X$ denoted by 
${p}_{X}=\{{p}_{X} (x)\}_{x \in \mathcal{X}}$.
The stationary DMS $\{X_t\}_{t=1}^\infty$ 
is specified with ${p}_{X}$.
Also, let $K$ be a random variable taken from 
the same finite set $\mathcal{X}$ representing the key 
used for encryption.
Similarly, let $\{K_{t}\}_{t=1}^\infty$ be a stationary 
DMS such that for each $t=1,2,\ldots$, $K_{t}$ 
takes values in the finite set
$\mathcal{X}$ and obeys the same distribution as that of $K$ denoted by 
${p}_{K}=\{{p}_{K} (k)\}_{k\in\mathcal{X}}$.
The stationary DMS $\{K_t\}_{t=1}^\infty$ 
is specified with ${p}_{K}$. 

\noindent{}\textit{Random Variables and Sequences: \ }
We write the sequence of random variables with length $n$ 
from the information source as follows:
${\rvcx}\defeq X_{1}X_{2}\cdots X_{n}$. 
Similarly, the strings with length 
$n$ of $\mathcal{X}^n$ are written as 
${\vcx}\defeq x_{1}x_{2}\cdots
x_{n}\in\mathcal{X}^n$. 
For ${\vcx}\in \mathcal{X}^n$, 
${p}_{{\lrvcx}}({\vcx})$ stands for the 
probability of the occurrence of 
${\vcx}$. 
When the information source is memoryless 
specified with ${p}_{X}$, we have 
the following equation holds:
$$
{p}_{{\lrvcx}}({\vcx})=\prod_{t=1}^n {p}_{X}(x_t).
$$
In this case we write ${p}_{{\lrvcx}}({\vcx})$
as ${p}_{X}^n({\vcx})$. Similar notations are 
used for other random variables and sequences.

\noindent{}\emph{Consensus and Notations: }
Without loss of generality, throughout this paper,
we assume that $\mathcal{X}$ is a finite field.
The notation $\oplus$ is used to denote the field 
addition operation, while the notation $\ominus$ 
is used to denote the field subtraction operation, i.e., 
$a\ominus b = a \oplus (-b)$ for any elements 
$a,b \in {\cal X}$. Throughout 
this paper all logarithms are 
taken to the base natural.

\subsection{Basic System Description}

In this subsection we explain the basic system setting and basic 
adversarial model we consider in this paper. First, let the information 
source and the key be generated independently by different parties $\Sgen$ 
and $\Kgen$ respectively. We further assume that the source 
is generated by $\Sgen$ and independent of the key. 

\begin{figure}[t]
\centering
\includegraphics[width=0.43\textwidth]{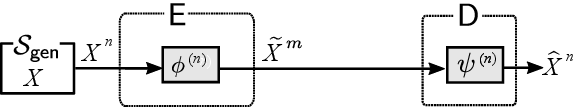}
\caption{Source coding without encryption.
\label{fig:mainA}}
\end{figure}

\noindent
\underline{\it Source coding without encryption:} \ The 
random source ${\rvcx}$ from $\Sgen$ be sent to node 
$\mathsf{E}$. Further settings of the system are 
described as follows. 
Those are also shown in Fig. \ref{fig:mainA}.
\begin{enumerate}
	\item \emph{Encoding Process:} \ 
        At the node $\mathsf{E}$, the encoder function 
        $\phi^{(n)}: {\cal X}^n $ $\to {\cal X}^{m}$ 
        observes ${\rvcx}$ to generate 
        $\tilde{X}^{m}=\phi^{(n)}({\rvcx})$. 
        Without loss of generality we may assume that 
        $\phi^{(n)}$ is {\it surjective}. 
	\item \emph{Transmission:} \ 
        Next, the encoded source $\tilde{X}^{m}$ is 
        sent to the destination $\D$ through a \emph{noiseless} 
        channel. 
	\item \emph{Decoding Process:} \ 
        In $\D$, the decoder function observes $\tilde{X}^{m}$
        to output ${\hrvcx}$, 
        using the one-to-one mapping $\psi^{(n)}$ defined by 
        $\psi^{(n)}:
             {\cal X}^{m} \to {\cal X}^n$. 
        Here we set 
        \begin{align*}
         \hrvcx \defeq & \psi^{(n)}(\tilde{X}^{m})
         = \psi^{(n)}\left(\phi^{(n)}(\rvcx)\right).
        \end{align*}
More concretely, the decoder outputs the unique pair 
$\hrvcx$ from $(\phi^{(n)})^{-1}(\tilde{X}^{m})$ 
in a proper manner. 
\end{enumerate}
For the above $(\phi^{(n)},\psi^{(n)})$, 
define the set $ \mathcal{D}^{(n)}$ of correct decoding by 
$
\mathcal{D}^{(n)} 
:=
\{\vcx \in \mathcal{X}^{n}:\psi^{(n)}(\phi^{(n)}(\vcx))=\vcx\}.
$
On $|{\cal D}^{(n)}|$, we have the following property. 
\begin{property}\label{pr:prOnDecSet} 
Under the conditions 1)-3) that we assume 
in the distributed source coding without encryption we have 
$|{\cal D}^{(n)}|=|{\cal X}^{m}|$. 
\end{property}

\newcommand{\ASSxx}{
\begin{IEEEproof}
We have the following:
\begin{align}
{\cal D}^{(n)}\MEq{a}& 
\{\vcxone =\psi^{(n)}(\tilde{x}^{m}):
\tilde{x}^{m} \in \phi^{(n)}({\cal X}^{n})\}
\notag\\
\MEq{b}&
\{ \vcx=\psi^{(n)}(\tilde{x}^{m}):
\tilde{x}^{m} \in {\cal X}^{m} \}.
\label{eqn:pSddxcc} 
\end{align}
Step (a) follows from that every pair 
$ \tilde{x}^{m} \in \phi^{(n)}({\cal X}^{n})$ uniquely 
determines $\vcx \in {\cal D}^{(n)}$.
Step (b) follows from that $\phi^{(n)}$ are surjective.
Since 
$\psi^{(n)}: {\cal X}^{m} \to {\cal X}^n$ is a one-to-one mapping 
and (\ref{eqn:pSddxcc}), 
we have $|{\cal D}^{(n)}|=|{\cal X}^{m}|.$
\end{IEEEproof}
}

Proof of Property \ref{pr:prOnDecSet} is given 
in Appendix \ref{apd:ProofPrOnDecSet}. 

\newcommand{\ProofPrOnDecSet}{
\subsection{
Proof of Property \ref{pr:prOnDecSet} 
}
\label{apd:ProofPrOnDecSet}
In this appendix we prove the property on 
the decoding set ${\cal D}^{(n)}$ stated 
in Property \ref{pr:prOnDecSet}. 

\begin{IEEEproof}[Proof of Property \ref{pr:prOnDecSet}] 
We have the following:
\begin{align}
{\cal D}^{(n)}\MEq{a}& 
\{\vcxone =\psi^{(n)}(\tilde{x}^{m}):
\tilde{x}^{m} \in \phi^{(n)}({\cal X}^{n})\}
\notag\\
\MEq{b}&
\{ \vcx=\psi^{(n)}(\tilde{x}^{m}):
\tilde{x}^{m} \in {\cal X}^{m} \}.
\label{eqn:pSddxcc} 
\end{align}
Step (a) follows from that every pair 
$ \tilde{x}^{m} \in \phi^{(n)}({\cal X}^{n})$ uniquely 
determines $\vcx \in {\cal D}^{(n)}$.
Step (b) follows from that $\phi^{(n)}$ are surjective.
Since 
$\psi^{(n)}: {\cal X}^{m} \to {\cal X}^n$ is a one-to-one mapping 
and (\ref{eqn:pSddxcc}), 
we have $|{\cal D}^{(n)}|=|{\cal X}^{m}|.$
\end{IEEEproof}
}

\begin{figure}[t]
\centering
\includegraphics[width=0.43 \textwidth]{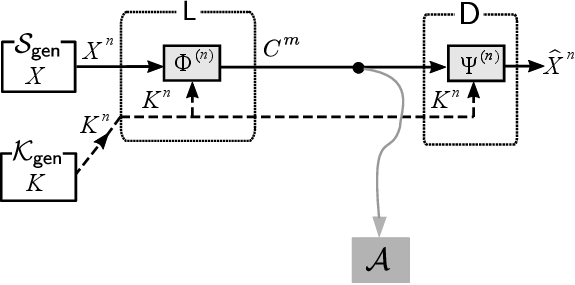}
\caption{Source coding with encryption.
\label{fig:main}}
\end{figure}

\noindent
\underline{\it Source coding with encryption:} \ 
The source ${\rvcx}$ from $\Sgen$ is sent to the node 
$\mathsf{L}$. The random key ${\rvck}$ from $\Kgen$, 
is sent to $\mathsf{L}$. Further settings of our system 
are described as follows. 
Those are also shown in Fig. \ref{fig:main}.
\begin{enumerate}
\item \emph{Source Processing:} At the node $\L$, ${\rvcx}$ 
is encrypted with the key ${\rvck}$ using the encryption function 
        $\Phi^{(n)}:{\cal X}^n \times {\cal X}^n$ 
       	$\to {\cal X}^{m}$. 
        The ciphertext $C^{m}$ of ${\rvcx}$ is given by 
        $C^{m}=\Phi^{(n)}({\rvck},{\rvcx})$. 
        On the encryption function $\Phi^{(n)}$, we 
        use the following notation:
        $$
         \Phi^{(n)}({\rvck}, {\rvcx})
        =\Phi^{(n)}_{{\lrvck}}({\rvcx})
        =\Phi^{(n)}_{{\lrvcx}}({\rvck}).
        $$  
        \item \emph{Transmission:} \ Next, the ciphertext 
        $C^{m}$ is sent to the destination $\D$ through 
        the \emph{public} communication channel. 
        Meanwhile, the key ${\rvck}$ is sent 
        to $\D$ through the \emph{private} communication channel.
        \item \emph{Sink Node Processing:} \ In $\D$, we decrypt 
        the ciphertext $\hrvcx$ from $C^{m}$ using 
        the key ${\rvck}$ through the corresponding 
        decryption procedure $\Psi^{(n)}$ defined by 
        $ \Psi^{(n)}: {\cal X}^n \times {\cal X}^{m} 
         \to {\cal X}^n$.
        Here we set 
        $ \hrvcx \defeq \Psi^{(n)}({\rvck},C^{m}).$ 
        More concretely, the decoder outputs the unique 
        $\hrvcx$ from $(\Phi_{\lrvck}^{(n)})^{-1}({C}^{m})$ 
        in a proper manner. On the decryption function 
        $\Psi^{(n)}$, we use the following notation:
        \begin{align*}
        & \Psi^{(n)}({\rvck},C^{m})
         =\Psi^{(n)}_{\lrvck}(C^{m})
         =\Psi^{(n)}_{C^{m}}({\rvck}).
        \end{align*}  
\end{enumerate}

Fix any $ \rvck=\vck \in \mathcal{X}^{n}$.
For this $\rvck$ and 
for $(\Phi^{(n)},\Psi^{(n)})$,  
we define the set $ \mathcal{D}^{(n)}_{\lvck}$ 
of correct decoding by 
\begin{align*}
\mathcal{D}^{(n)}_{\lvck}
& := \{\vcx \in \mathcal{X}^{n}:
\Psi_{\lvck}^{(n)}( \Phi_{\lvck}^{(n)} (\vcx))= \vcx \}.
\end{align*}
We require that the cryptosystem 
$
(\Phi^{(n)},\Psi^{(n)})
$
must satisfy the following condition.

{\it Condition:} 
For each distributed source encryption system 
$(\Phi^{(n)},\Psi^{(n)})$,  
there exists a source coding system 
$(\phi^{(n)},\psi^{(n)})$ such that  
for any $\vck \in \mathcal{X}^{n}$ and for any 
$\vck \in \mathcal{X}^{n}$, 
\begin{align*}
& \Psi_{\lvck}^{(n)}(\Phi_{\lvck}^{(n)}(\vcx)) 
  =\psi^{(n)}(\phi^{(n)}(\vcx)). 
\end{align*}
The above condition implies that 
${\cal D}^{(n)}={\cal D}^{(n)}_{\lvck}, \forall \vck \in {\cal X}^n.$
We have the following properties on ${\cal D}^{(n)}$. 
\begin{property}
\label{pr:prOne}
$\quad$
\begin{itemize}
\item[a)] 
If $\vcx, {\vcy} \in {\cal D}^{(n)}$ and 
   $\vcx \neq {\vcy}$, then 
$\Phi^{(n)}_{\lvck}(\vcx) \neq \Phi^{(n)}_{\lvck}({\vcy}).$
\item[b)] 
    $\forall \vck$ and $\forall c^{m}$, 
$\exists \vcx$ $\in {\cal D}^{(n)}$ such that  
$
\Phi^{(n)}_{\lvck}(\vcx)=c^{m}.
$
\end{itemize}
\end{property}

Proof of Property \ref{pr:prOne} is given 
in Appendix \ref{apd:ProofPrOne}. 
\newcommand{\ProofPrOne}{
\subsection{
Proof of Property \ref{pr:prOne}
}\label{apd:ProofPrOne}

We first prove the part a) and next prove the part b).   
\begin{IEEEproof}[Proof of Property \ref{pr:prOne} part a)] 
Under $\vcx, \vcy \in {\cal D}^{(n)}$ and 
$\vcx \neq$  $\vcy$, we assume that 
\beq
 \Phi^{(n)}_{\lvck}({\vcx})=\Phi^{(n)}_{\lvck}({\vcy}).
\label{eqn:Assum} 
\eeq
Then  we have the following 
\begin{align}
&\vcx \MEq{a}
\psi^{(n)}(\phi^{(n)}(\vck) \MEq{b} \Psi^{(n)}_{\lvck}
(\Phi^{(n)}_{\lvck}({\vcx}))  
\notag\\
& \MEq{c}
\Psi^{(n)}_{\lvck}(\Phi^{(n)}_{\lvck}(\vcy))
\MEq{d}
 \psi^{(n)}(\phi^{(n)}(\vcy))\MEq{e}{\vcy}.
\label{eqn:SdCCv}
\end{align}
Steps (a) and (e) follow from the definition of 
${\cal D}^{(n)}$. Step (c) follows from (\ref{eqn:Assum}).
Steps (b) and (d) follow from the relationship between 
$(\phi^{(n)},\psi^{(n)})$ and 
$(\Phi^{(n)}_{\lvck}, \Psi^{(n)}_{\lvck}).$
The equality (\ref{eqn:SdCCv}) contradics the first assumption.
Hence we must have Property \ref{pr:prOne} part a).
\end{IEEEproof}
\begin{IEEEproof}[Proof of Property \ref{pr:prOne} part b)] 
We assume that $\exists \vck$ and 
$\exists c^{m}$ such that $\forall \vcx \in {\cal D}^{(n)}$, 
$\Phi^{(n)}_{\lvck}(\vcx)$ $\neq$ $c^{m}$. Set 
$$
{\cal B} \defeq \left\{ \Phi^{(n)}_{\lvck}(\vcx): 
\vcx \in {\cal D}^{(n)}
\right\}.
$$  
Then by the above assumption we have
\begin{align}
&{\cal B} \subseteq {\cal X}^{m} -\left\{c^{m}\right\}.
\label{eqn:SdCCvpp}
\end{align} 
On the other hand we have 
\begin{align*}
   \Psi^{(n)}_{\lvck}({\cal B})
&= \left\{
   \Psi^{(n)}_{\lvck}( \Phi^{(n)}_{\lvck}(\vcx)):
     \vcx \in {\cal D}^{(n)}
\right\}
\\
&= \left\{
   \psi^{(n)}(\phi^{(n)}(\vcx)): \vcx \in {\cal D}^{(n)}
\right\}
={\cal D}^{(n)},
\end{align*}
which together with that 
$\Psi_{\lvck}:$ ${\cal X}^{m}$ $\to$ ${\cal X}^{n}$
is a one-to-one mapping yields that 
\begin{align*}
& |{\cal B}|=|\Psi^{(n)}_{\lvck}({\cal B})|
 =|{\cal D}^{(n)}|=|{\cal X}^{m}|.
\end{align*} 
The above equality contradicts (\ref{eqn:SdCCvpp}). 
Hence we must have that $\forall \vck$, $\forall c^{m}$,
$\exists \vcx \in {\cal D}^{(n)}$ such that 
$\Phi^{(n)}_{\lvck}(\vcx)=c^{m}$.
\end{IEEEproof}
}
\begin{figure}[t]
\centering
\includegraphics[width=0.43\textwidth]{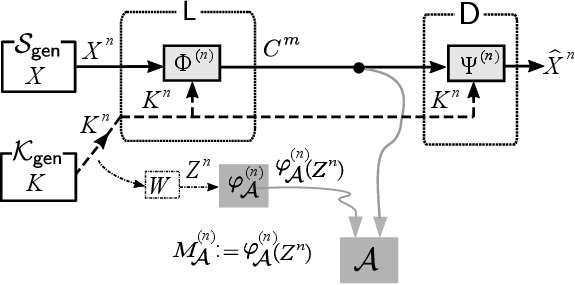}
\vspace*{-2mm}
\caption{Side-channel attacks to the source coding with encryption.
\label{fig:mainZ}}
\vspace*{-2mm}
\end{figure}

\noindent
\underline{\it Side-Channel Attacks by Eavesdropper Adversary:} 
An 
\emph{adversary} $\A$ eavesdrops the public 
communication channel in the system. The adversary $\A$ 
also uses a side information obtained by side-channel attacks. 
Let ${\cal Z}$ be a finite set and let $W:{\cal X}\to {\cal Z}$ 
be a noisy channel. Let $Z$ be a channel output from $W$ for the 
input random variable $K$. We consider the discrete memoryless 
channel specified with $W$. Let $\rvcz \in {\cal Z}^n$ be 
a random variable obtained as the channel output by connecting 
$\rvck \in {\cal X}^n$ to the input of channel. 
We write a 
conditional distribution on $\rvcz$ given $\rvck$ as 
$$
W^n=
\left\{W^n(\vcz|\vck)\right
\}_{(\lvck,\lvcz)\in {\cal K}^n \times {\cal Z}^n}.
$$
On the above output $\rvcz$ of $W^n$ for the input $\rvck$, 
we assume the followings.
\begin{itemize}
\item
The three random variables $X$, $K$ and $Z$, satisfy 
$X \perp (K,Z)$, which implies that $X^n \perp (K^n,Z^n)$.
\item $W$ is given 
in the system and the adversary ${\cal A}$ can not control $W$. 
\item By side-channel attacks, the adversary ${\cal A}$ 
can access $Z^n$. 
\end{itemize}
We next formulate side information the adversary ${\cal A}$ 
obtains by side-channel attacks. For each $n=1,2,\cdots$, 
let $\varphi_{\cal A}^{(n)}:{\cal Z}^n 
\to {\cal M}_{\cal A}^{(n)}$ be an encoder function. 
Set $M_{\cal A}^{(n)} \defeq \varphi_{\cal A}^{(n)}(Z^n) 
\in {\cal M}_{\cal A}^{(n)}$. We assume that 
$||\varphi_{\cal A}^{(n)}||=|{\cal M}_{\cal A}^{(n)}|$ must satisfy 
the rate constraint 
$
||\varphi_{\cal A}^{(n)}||\leq {\rm e}^{n R_{\cal A}}.
$ 

\newcommand{\DeLoNe}{
Furthermore, set 
$\varphi_{\cal A} \defeq \{\varphi_{\cal A}^{(n)}\}_{n=1,2,\cdots}.$ 
Let 
$$ 
R_{\cal A}^{(n)}\defeq 
\frac{1}{n} \log ||\varphi_{\cal A}||
=\frac{1}{n} \log |{\cal M}_{\cal A}^{(n)}|
$$
be a rate of the encoder function $\varphi_{\cal A}^{(n)}$. 
In the following arguments all logarithms are 
taken to the base natural. 
For $R_{\cal A}>0$, we set 
${\cal F}_{\cal A}^{(n)}(R_{\cal A})
\defeq \{ \varphi_{\cal A}^{(n)}: R_{\cal A}^{(n)} \leq R_{\cal A}\}.$ 
}
\subsection{
Security Criterion and Problem Set Up}

In this subsection, we propose a new security criterion. 
We first state a lemma having a close connection with 
the new security criterion. This lemma is shown below.  
\begin{lemma}\label{lem:LemOne} 
$\forall (c^{m},a) \in {\cal X}^{m} 
\times {\cal M}_{\cal A}^{(n)}$, we have 
\begin{align*}
\sum_{\lvcx \in {\cal D}^{(n)}}
p_{C^m|M_{\cal A}^{(n)}\lrvcx} (c^{m}|a, \vcx)=1.
\end{align*}
\end{lemma}

Lemma \ref{lem:LemOne} can easily be proved by 
Property \ref{pr:prOne}. The detail 
is found in Appendix \ref{apd:ProofLemOne}. 
This lemma can be regarded as an extension of the 
Birkhoff-von Neumann theorem \cite{iwamoto:11}.
\newcommand{\ProofLemOne}{
\subsection{Proof of Lemma \ref{lem:LemOne}}
\label{apd:ProofLemOne}
In this appendix we prove Lemma \ref{lem:LemOne}. 
For $\vcx \in {\cal X}^n $, we set 
\begin{align*}
&{\cal A}_{\lvcx}(c^{m})
 \defeq \left\{ \vck: \Phi^{(n)}_{\lvcx}({\vck})=c^{m}\right\}.
\end{align*}
\begin{IEEEproof}[Proof of Lemma \ref{lem:LemOne}]
Property \ref{pr:prOne} part a) 
implies that 
\begin{align} & 
{\cal A}_{\lvcx}(c^{m}) \cap {\cal A}_{\lvcy}(c^{m})=\emptyset
\mbox{ for } \vcx \neq {\vcy} \in {\cal D}^{(n)}. 
\label{eqn:Serrq}
\end{align}
Furthermore, Property \ref{pr:prOne} part b) implies that  
\begin{align}
 \bigcup_{ \lvcx \in {\cal D}^{(n)} }{\cal A}_{\lvcx}(c^{m})
={\cal X}^{n}. \label{eqn:SerrqB}
\end{align}
From (\ref{eqn:ddrtrq}), for each 
$(c^m,a) \in {\cal X}^m \times {\cal M}_{\cal A}^{(n)}$, 
we have the following chain of equalities:
\begin{align*}
& \sum_{\lvcx \in {\cal D}^{(n)}}
p_{C^m|M_{\cal A}^{(n)}\lrvcx} (c^{m}|a, \vcx)
\\
&\MEq{a}\Pr \left\{\rvck
  \in \bigcup_{ \lvcx \in {\cal D}^{(n)}} {\cal A}_{\lvcx}(c^{m})
  \Hugepr {\cal M}_{\cal A}^{(n)}=a \right\} \MEq{b}1. 
\end{align*}
Step (a) follows from (\ref{eqn:Serrq}). 
Step (b) follows from (\ref{eqn:SerrqB}).
\end{IEEEproof} 
}

In the following arguments all logarithms are taken to the base natural.  
The adversary ${\cal A}$ tries to estimate ${\rvcx} \in \mathcal{X}^n$ 
from
$(C^{m},$ $M_{\cal A}^{(n)})$.
Note that since $X^n \perp (K^n,Z^n)$, we have 
$X^n \perp (K^n,$ $M_{\cal A}^{(n)})$. 
The mutual information (MI) between $\rvcx$ and 
$(C^{m},M_{\cal A}^{(n)})$ 
denoted by 
$$
\Delta_{\rm MI}^{(n)}
\defeq I(C^{m}M_{\cal A}^{(n)};\rvcx)
=I(C^{m};\rvcx|M_{\cal A}^{(n)})
$$
indicates a leakage of information on $\rvcx$ 
from $(C^{m},$ $M_{\cal A}^{(n)})$. 
In this sense it seems to be quite natural 
to adopt the mutual information $\Delta_{\rm MI}^{(n)}$ 
as a security criterion. 
On the other hand, 
directly using $\Delta_{\rm MI}^{(n)}$ as a security criterion 
of the cyptosystem has some problem that this value 
depends on the statistical property of $\rvcx$. 
In this paper we propose a new security criterion, which is based on
$\Delta_{\rm MI}^{(n)}$ but overcomes the above problem. 
\begin{definition} 
Let $\overline{X}^n $ be an arbitrary random variable 
taking values in ${\cal X}^n$. 
Set $\overline{C}^m=\Phi^{(n)}(\rvck,\overline{X}^n)$.
The maximum mutual information criterion denoted 
by $\Delta_{\rm max-MI}^{(n)}$ is as follows. 
\begin{align*}
&\Delta_{{\rm max-MI}}^{(n)} \defeq \max_{p_{\overline{X}^n} 
\in {\cal P}({\cal X}^n)}
I(\overline{C}^{m} ; \overline{X}^n| M_{\cal A}^{(n)}).
\end{align*}
\end{definition} 

By definition it is obvious that 
$\Delta_{\rm MI}^{(n)} \leq {\Delta}_{{\rm max-MI}}^{(n)}$.
We have the following proposition on 
$\Delta_{{\rm max-MI}}^{(n)}$: 
\begin{proposition}\label{pro:ProOneA}
$\quad$
\begin{itemize}
\item[a)] If 
$ \Delta_{\rm MI}^{(n)}=I(C^{m};\rvcx | M_{\cal A}^{(n)})=0,$ 
then, we have $\Delta_{{\rm max-MI}}^{(n)}=0$. 
This implies that $\Delta_{{\rm max-MI}}^{(n)}$ is valid as a 
measure of information leakage. 
\item[b)] We have the following.
\begin{align*}
& {\Delta}_{{\rm max-MI}}^{(n)} \geq m\log{|{\cal X}|}- H(K^{n}|M_{\cal A}^{(n)}).
\end{align*}
\end{itemize}
\end{proposition}

Proof of Proposition \ref{pro:ProOneA} is given in 
Appendix \ref{apd:ProofProOneA}.
\newcommand{\ProofProOneA}{
\subsection{Proof of Proposition \ref{pro:ProOneA}}
\label{apd:ProofProOneA}

In this appendix we prove Proposition \ref{pro:ProOneA}. 
We first give some preliminary results nessary for the proof.
By the definition of ${\cal A}_{\lvcx}(c^{m})$, 
we have that for each $(c^m,a,\vcx) \in {\cal X}^m \times 
{\cal M}_{\cal A}^{(n)} \times {\cal X}^n$
\begin{align}
&p_{C^{m}|M_{\cal A}^{(n)}\lrvcx}(c^m|a,\vcx)
\nonumber\\
&={\rm Pr}\left\{\rvck \in {\cal A}_{\lvcx}(c^{m})
\Bigl| M_{\cal A}^{(n)}=a, \rvcx=\vcx \right\}
\nonumber\\
&\MEq{a} {\rm Pr}\left\{\rvck \in {\cal A}_{\lvcx}(c^{m})
\Bigl| M_{\cal A}^{(n)}=a \right\}.
\label{eqn:ddrtrq}
\end{align}
Step (a) follows from $(\rvck, M_{\cal A}^{(n)}) \perp \rvcx$.
We can see from (\ref{eqn:ddrtrq}) that 
for each $\vcx \in {\cal X}^n$, 
the component $p_{C^m|M_{\cal A}^{(n)}\lrvcx}(c^m$ $|a,\vcx)$ 
of the stochastic matrix 
\begin{align*}
& p_{C^m|M_{\cal A}^{(n)}\lrvcx}(\cdot|\cdot,\vcx) 
\\
&= \left\{ 
p_{C^m|M_{\cal A}^{(n)}\lrvcx}(c^m|a,\vcx)
   \right\}_{ 
(c^m,a) \in {\cal X}^m \times {\cal M}_{\cal A}^{(n)}
}\end{align*}
can be written as 
\begin{align*}
& p_{C^m| M_{\cal A}^{(n)} \lrvcx}(c^m|a,\vcx)
=\Gamma_{\lrvck M_{\cal A}^{(n)}, \vcx}(c^m|a). 
\end{align*}
Furthermore, the quantity 
\begin{align*}
&\Gamma_{\lrvck M_{\cal A}^{(n)},\vcx}
\defeq \left\{ 
\Gamma_{\lrvck M_{\cal A}^{(n)},\vcx}(c^m|a)
\right \}_{(c^m,a) 
\in {\cal X}^m \times {\cal M}_{\cal A}^{(n)}}
\end{align*}
can be regarded as a stochastic matrix 
indexed by $\vcx \in {\cal X}^n$.
Here the random pair 
$\rvck M_{\cal A}^{(n)}$ appearing in $\Gamma_{\lrvck M_{\cal A}^{(n)},\vcx}$ 
stands for that the randomness of the stochastic matrix is from that of 
$(\rvck, M_{\cal A}^{(n)})$. 

\begin{IEEEproof}[Proof of Proposition \ref{pro:ProOneA}] 

We first prove the part a). Using the quntities 
$$ 
\Gamma_{\lrvck M_{\cal A}^{(n)},\lvcx}(c^m|a), (\vcx, c^m,a) 
   \in {\cal X}^n 
\times {\cal X}^m 
\times {\cal M}_{\cal A}^{(n)},
$$
components $p_{C^m|M_{\cal A}^{(n)}\lrvcx}(c^m|a)$ 
of the stochastic matrix $p_{C^m|M_{\cal A}^{(n)}}$ can 
be computed as 
\begin{align*}
p_{C^m|M_{\cal A}^{(n)}}(c^m|a)=
\sum_{\lvcx} p_{X^n}(\vcx)
\Gamma_{\lrvck M_{\cal A}^{(n)},\lvcx}(c^m|a). 
\end{align*}
Set 
\begin{align*}
&\Gamma_{\lrvck M_{\cal A}^{(n)}}^{(p_{X^n})}(c^m|a)
=\sum_{\lvcx} p_{X^n}(\vcx)
 \Gamma_{\lrvck M_{\cal A}^{(n)},\lvcx}(c^m|a)
\\
&= p_{C^m|M_{\cal A}^{(n)}}(c^m|a).
\end{align*}
Furthermore, set 
\begin{align*}
& \Gamma^{(p_{X^n})}_{\lrvck M_{\cal A}^{(n)}}
\defeq \left\{
\Gamma_{\lrvck M_{\cal A}^{(n)}}^{(p_{X^n})}(c^m|a)
\right\}_{(c^m,a) 
\in {\cal X}^m \times {\cal M}_{\cal A}^{(n)} }
\\
&=p_{C^m|M_{\cal A}^{(n)}}.
\end{align*}
Using 
$\Gamma_{\lrvck M_{\cal A}^{(n)}, 
\lvcx}, 
 \vcx \in {\cal X}^n$ and 
$\Gamma_{\lrvck M_{\cal A}^{(n)}}^{(p_{X^n})}$,    
we compute $\Delta_{\rm MI}^{(n)}$ to obtain 
\begin{align}
&\Delta_{\rm MI}^{(n)}=I(C^{m}; \rvcx | M_{\cal A}^{(n)})
=\sum_{\lvcx\in {\cal X}^n}p_{\lrvcx}(\vcx)
\notag\\
&\quad \qquad \times D\left(
\Gamma_{\lrvck M_{\cal A}^{(n)},\lvcx} \Big|\Big| 
\Gamma^{(p_{X^n})}_{\lrvck M_{\cal A}^{(n)}}
\Big| p_{M_{\cal A}^{(n)}} \right).
\label{eqn:SDDxx}
\end{align}
We note that since $\rvcx$ is from the discrete memolyless 
source specified with $p_{X}$, we have that 
\begin{align}
&p_{\lrvcx}(\vcx) = \prod_{t=1}^n p_{X}(x_t)>0, 
\forall \vcx \in {\cal X}^n.
\label{eqn:Sffxx}
\end{align}
Now we suppose that $\Delta_{\rm MI}^{(n)}=0$. 
Then from (\ref{eqn:SDDxx}) and (\ref{eqn:Sffxx}), we have 
\begin{align}
& \Gamma_{\lrvck M_{\cal A}^{(n)},\vcx }
= \Gamma_{\lrvck M_{\cal A}^{(n)},0^n  }
= \Gamma^{(p_{X^n})}_{\lrvck M_{\cal A}^{(n)}}, 
\forall \vcx \in {\cal X}^n. 
\label{eqn:ssSDDxppx}
\end{align}
Let $\overline{X}_{\rm opt}^n$ 
be the optimal random variable, the distribution 
$p_{\overline{X}_{\rm opt}^n}$ of which 
attains the maximum in the definition of 
$\Delta_{{\rm max-MI}}^{(n)}$. Set 
$\overline{C}^m_{\rm opt}
=\Phi^{(n)}(K^n,\overline{X}_{\rm opt}^n)$. 
By definition we have 
$\Delta_{{\rm max-MI}}^{(n)}
=I(\overline{C}_{\rm opt}^m; 
\overline{X}_{\rm opt}^n |M_{\cal A}^{(n)})$. 
Using (\ref{eqn:ssSDDxppx}), we compute 
$\Gamma^{(p_{\overline{X}_{\rm opt}^n })}_{\lrvck M_{\cal A}^{(n)}}(c^m|a)$, 
$(c^m,a) \in {\cal X}^m \times {\cal M}_{\cal A}^{(n)}$ to obtain
\begin{align*}
&\Gamma_{\lrvck M_{\cal A}^{(n)}}^{(p_{\overline{X}_{\rm opt}^n})}(c^m|a)
=\sum_{\lvcx} p_{\overline{X}_{\rm opt}^n}(\vcx)
 \Gamma_{\lrvck M_{\cal A}^{(n)},0^n}(c^m|a)
\notag\\
&= \Gamma_{\lrvck M_{\cal A}^{(n)},0^n}(c^m|a).
\end{align*}
Hence we have
\beq
\Gamma^{(p_{\overline{X}_{\rm opt}^n })}_{\lrvck M_{\cal A}^{(n)}}
= \Gamma_{\lrvck M_{\cal A}^{(n)},0^n}=
\Gamma_{\lrvck M_{\cal A}^{(n)},\lvcx},
\forall \vcx \in {\cal X}^n. 
\label{eqn:jjDDxppx}
\eeq 
From (\ref{eqn:jjDDxppx}), we have
\begin{align*}
&\Delta_{{\rm max-MI}}^{(n)}
=I(\overline{C}_{\rm opt}^m;\overline{X}_{\rm opt}^n | M_{\cal A}^{(n)})
 =\sum_{\lvcx\in {\cal X}^n}p_{ \overline{X}_{\rm opt}^n}(\vcx)
\\
& \quad \qquad \times 
D\left(
\Gamma_{\lrvck M_{\cal A}^{(n)},\lvcx} \Big|\Big| 
\Gamma^{(p_{\overline{X}_{\rm opt}^n })}_{\lrvck M_{\cal A}^{(n)}}
\Big| p_{M_{\cal A}^{(n)}} 
\right)=0.
\end{align*}
We next prove the part b).
\begin{figure}[t]
\centering
\includegraphics[width=0.28\textwidth]{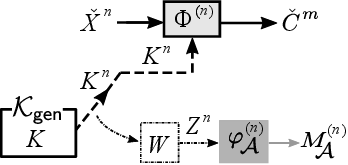}
\caption{$\crvcx$, $\crvcc$ and $M_{\cal A}^{(n)}$.
\label{fig:CheckRV}}
\end{figure}
Let $\crvcx$ be a uniformly distributed random variable 
over ${\cal D}^{(n)}$. Set $\check{C}^{m} 
\defeq$ $\Phi_{\lrvck}(\crvcx)$.
The three random variables $\crvcx$, 
$\check{C}^{m}$, and $M_{\cal A}^{(n)}$ are shown 
in Fig. \ref{fig:CheckRV}. 
$\check{C}^{m}$ is the uniformly distributed random variable
 over ${\cal X}^{m}$ and independent of ${M}_{\cal A}^{(n)}$.
In fact for each 
$(c^m,a) \in {\cal X}^m \times {\cal M}_{\cal A}^{(n)}$, 
we have the following chain of equalities:
\begin{align}
& p_{\check{C}^{(m)}| M_{\cal A}^{(n)} }(c^{m}|a)
=\sum_{\scs \lvcx {\scs \in {\cal D}^{(n)}}}
\hspace*{-2mm}
p_{\check{C}^{(m)}| M_{\cal A}^{(n)} \lcrvcx}
(c^{m}|a,\vcx) \cdot \frac{1}{|{\cal X}^{m}|}
\notag\\
&=\frac{1}{|{\cal X}^{m}|}
 \sum_{\scs \lvcx 
       {\scs \in {\cal D}^{(n)}}}
\Gamma_{\lrvck M_{\cal A}^{(n)},\lvcx}
(c^{m}|a)
\MEq{a}\frac{1}{|{\cal X}^{m}|}.
\label{eqn:SDfffp}
\end{align}
Step (a) follows from Lemma \ref{lem:LemOne}.
Since we have (\ref{eqn:SDfffp}) for every 
$(c^m,a) \in {\cal X}^m \times {\cal M}_{\cal A}^{(n)}$, 
we have that $\check{C}^m$ is the uniformly distributed 
random variable over ${\cal X}^m$ and independent of 
$M_{\cal A}^{(n)}$. We have the following chain of inequalities:
\begin{align*}
& \Delta_{{\rm max-MI}}^{(n)}\geq I(\check{C}^m ; \crvcx | M_{\cal A}^{(n)})
\nonumber\\ 
&= H(\check{C}^m|M_{\cal A}^{(n)})-
   H(\check{C}^m|M_{\cal A}^{(n)},\crvcx)
\nonumber\\ 
&\MEq{a} m \log |{\cal X}|-H(\check{C}^m|M_{\cal A}^{(n)},\crvcx).
\\
&=m \log |{\cal X}|
-H(\Phi^{(n)}({\rvck},\crvcx)| M_{\cal A}^{(n)}, \crvcx)
\\
& \MGeq{b} m\log{|{\cal X}|}-H(K^n|M_{\cal A}^{(n)},\crvcx )
\\
&=m\log{|{\cal X}|}-H(K^n|M_{\cal A}^{(n)}).
\end{align*}
Step (a) follows from Lemma \ref{lem:LemOne} part c). 
Step (b) follows from the data processing inequality.
\end{IEEEproof}
\noindent
}
\begin{remark}
The part a) in the above proposition is quite essential. If we have 
a security criterion $\widehat{\Delta}^{(n)}$ not satisfying this 
condition, it may happen that 
$
\Delta_{\rm MI}^{(n)}=I(C^{m};\rvcx | M_{\cal A}^{(n)})=0, 
$ 
but $\widehat{\Delta}^{(n)}>0$. Such $\widehat{\Delta}^{(n)}$ 
is invalid for the security criterion. 
\end{remark}
\begin{remark}
The property stated in the part b) is a key important property 
of $\Delta_{\rm max-MI}^{(n)}$, which plays an important role 
in establishing the strong converse theorem. Lemma \ref{lem:LemOne} 
is a key result for the proof of the part b).
\end{remark}

\noindent
\underline{\it Defining Reliability and Security:} 
The decoding process is successful if $\hrvcx=\rvcx$ holds.
Hence the decoding error probability 
is given by  
\begin{align*}
& \Pr[\Psi^{(n)}(\rvck,\phi^{(n)}({\rvck, \rvcx}))\neq {\rvcxone}]
\\
&= \Pr[\Psi^{(n)}_{\lrvck}
   (\Phi_{\lrvck}^{(n)}(\rvcx))\neq {\rvcxone} ]
\\
&= \Pr[\psi^{(n)}(\phi^{(n)}(\rvcxone)) \neq \rvcxone]
=\Pr[\rvcxone \notin {\cal D}^{(n)}].
\end{align*}
Since the above quantity depends only on 
$(\phi^{(n)},\psi^{(n)})$ and ${p}_{X}^n$,
we wirte the error probability $p_{{\rm e}}$ of decoding as 
\begin{align*}
p_{{\rm e}}=&p_{{\rm e}}(\phi^{(n)},\psi^{(n)}|{p}_{X}^n)
\defeq \Pr[\rvcxone \notin {\cal D}^{(n)}].
\end{align*}
Since $\Delta_{{\rm max-MI}}^{(n)}$ depends only on 
$(\Phi^{(n)},\varphi_{\cal A}^{(n)})$ 
and ${p}_{KZ}^n$, we write this quantity as 
as 
$
\Delta_{{\rm max-MI}}^{(n)}$ $=\Delta_{{\rm max-MI}}^{(n)}
(\Phi^{(n)},\varphi_{\cal A}^{(n)}|$ ${p}_{KZ}^n).
$
Define  
\begin{align*}
&\Delta_{{\cal A}}^{(n)}(\Phi^{(n)},R_{\cal A}|{p}_{KZ}^n)
\\
&\defeq \max_{\varphi_{\cal A}^{(n)}}
\left\{
\Delta_{{\rm max-MI}}^{(n)}(\Phi^{(n)}, \varphi_{\cal A}^{(n)}|{p}_{KZ}^n):
||\varphi_{\cal A}^{(n)}||\leq {\rm e}^{nR_{\cal A}}
\right\}.
\end{align*}

\begin{definition} We fix some positive constant $\varepsilon_0$. 
        For a fixed pair $(\varepsilon, \delta) 
        \in [0,\varepsilon_0] \times (0,1)$, a quantity $R$ 
        is $(\varepsilon,\delta)$-admissible 
        under $R_{\cal A}$ $>0$
        for the system $\mathsf{Sys}$ if 
        $\exists \{(\Phi^{(n)},$ $\Psi^{(n)})\}_{n \geq 1}$
        such that $\forall \gamma >0$,
        $\exists n_0=n_0(\gamma) \in \mathbb{N}$, 
	$\forall n\geq n_0$, 
	\begin{align*}
        &\frac{1}{n} \log |{\cal X}^{m}| 
         = \frac{m}{n} \log |{\cal X}| \in 
         \left[R-\gamma, R+ \gamma \right],
\\  				
& p_{{\rm e}}(\phi^{(n)},
              \psi^{(n)}|{p}_{X}^n)\leq \delta,\mbox{ and }
\Delta_{{\cal A}}^{(n)}(\Phi^{(n)},R_{\cal A}|{p}_{KZ}^n) 
\leq \varepsilon.
\end{align*}
\end{definition}

\begin{definition}{\bf (Reliable and Secure Rate Region)}
        Let $\mathcal{R}_{\mathsf{Sys}}(\varepsilon,$ $\delta|
        {p}_{X},$ ${p}_{KZ})$
	denote the set of all $(R_{\cal A},R)$ such that
        $R$ is $(\varepsilon,\delta)$-admissible under $R_{\cal A}$. 
        Furthermore, set 
        $$
         \mathcal{R}_{\mathsf{Sys}}
          (p_{X},p_{KZ}) \defeq  
        \bigcap_{\scs (\varepsilon, \delta) \in (0,\varepsilon_0] 
                 {\scs \times (0,1)
                 }
         }
\mathcal{R}_{\mathsf{Sys}}(\varepsilon,\delta|p_{X},p_{KZ}).
$$
We call $\mathcal{R}(p_{X},p_{KZ})$ 
the \emph{\bf reliable and secure rate} region. 
\end{definition}

\section{Direct Coding Theorem 
}

In this section we derive an explicit inner bound of 
$\mathcal{R}_{\mathsf{Sys}}(\varepsilon,\delta|p_{X},p_{KZ})$.
To derive this result we use our previous result \cite{santosoOh:19}. 
A condition for reliable transmission is an immediate consequence 
from the direct coding theorem for single discrete memoryless sources.
We derive a sufficient condition for secure transmission 
under the security criterion measured by $\Delta^{(n)}_{\rm max-MI}$.


\subsection{
Coding Scheme, Reliability and Security Analysis 
}

\begin{figure}[t] 
	\centering 
	\includegraphics[width=0.45\textwidth]
	{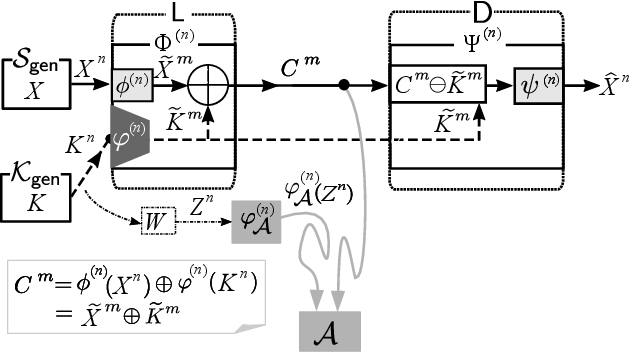}
\vspace*{-2mm}
	\caption{Proposed construction of 
        $(\Phi^{(n)},\Psi^{(n)})$.} 
        \label{fig:solution}%
\vspace*{-2mm}
\end{figure}

Our coding scheme is illustrated in Fig. \ref{fig:solution}. 
In this coding scheme we assume the following rate constraint: 
\begin{align}
& (1/n)\log |{\cal X}^{m}|= (m/n) \log |{\cal X}|
\in \left[R-(1/n), R\right].
\label{eqn:RateCond}
\end{align}
In the coding scheme in Fig. \ref{fig:solution}, 
we first provide a universal code construction of 
$\{(\phi^{(n)},\psi^{(n)})\}_{n\geq 1}$ deriving 
an exponential upper bound of 
$p_{\rm e}(\phi^{(n)},\psi^{(n)}|p_{X}^n)$.  
We next state a concreate construction of 
$\varphi^{(n)}$ in Fig. \ref{fig:solution}.
Based on  $(\phi^{(n)},\psi^{(n)})$ and 
$\varphi^{(n)}$, we construct $(\Phi^{(n)},\Psi^{(n)})$ in 
Fig. \ref{fig:solution}. We further provide some preliminary 
observation on an upper bound of 
$\Delta_{{\rm max-MI}}^{(n)}(\Phi^{(n)}$, 
$\varphi_{\cal A}^{(n)} |{p}_{KZ}^n)$. 

\noindent 
\underline{\it Universal Code Construction of 
$\{(\phi^{(n)}, \psi^{(n)})\}_{n \geq 1}$:} 
Let $\overline{X}$ be an arbitrary random variable
over $\mathcal{X}$ and has a probability distribution 
$p_{\overline{X}}$. Let $\mathcal{P}(\mathcal{\cal X})$ 
denote the set of all probability distributions on 
$\mathcal{X}$. Fix $\gamma>0$, arbitrary. For $ R \geq 0$ 
and $p_{X} \in $ $\mathcal{P}(\mathcal{\cal X})$, we define 
the following function:
\begin{align*}
	E_{\gamma}(R|p_{X}) &:{=}
	\min_{\scs p_{\overline{X}} \in 
	\mathcal{P}(\mathcal{\cal X}):
        \atop{\scs 
        R -\gamma \leq H(\overline{X})}}
		D(p_{\overline{X}}||p_X)\}.
\end{align*}
Set $\delta_{n} \defeq (1/n)\{|{\cal X}|\log(n+1)+1\}$.
Note that $\delta_{n} \to 0$ as  $ n\to \infty$. 
Let $n_0=n_0(\gamma)$ be the minimum integer such that we 
have $\delta_{ n} \leq \gamma$ for $n\geq n_0(\gamma)$. 
Then we have the following proposition.
\begin{proposition}\label{pro:ProCoding}
$\forall \gamma >0$, $\exists \{(\phi^{(n)}, 
                                 \psi^{(n)})\}_{n\geq 1}$ 
satisfying (\ref{eqn:RateCond}), such that 
$\forall p_X$ with $R > H(X)$ and 
$\forall n \geq n_0(\gamma)$, 
\begin{align}
& p_{\rm e}(\phi^{(n)},\psi^{(n)}|p_{X}^n) \leq 
(n+1)^{|{\cal X}|} {\ExP}^{-n E_{\gamma} (R|p_{X})}.
\label{eqn:mainZxxB}
\end{align}
\end{proposition}

Proposition \ref{pro:ProCoding} is a well known result on 
the universal coding for discrete memoryless sources. 
We omit the proof, which is found in \cite{HanKingo}.

\noindent
\underline{\it Affine Encoder as Privacy Amplifier:} 
Let $A$ is a matrix with $n$ rows and $m$ columns. Entries of 
$A$ are from ${\cal X}$. Let $b^{m}\in \mathcal{X}^{m}$. 
Define the mapping $\varphi^{(n)}: {\cal X}^n \to {\cal X}^{m}$
by 
\begin{align}
\varphi^{(n)}({\vck}):=&{\vck} A \oplus b^{m} 
\quad \mbox{ for }{\vck} \in \mathcal{X}^n.
\label{eq:homomorphic}
\end{align}
The mapping $\varphi^{(n)}$ is called the affine mapping. 

\noindent
\underline{\it Description of Proposed Procedure:} 
We describe out construction of $(\Phi^{(n)},\Psi^{(n)})$ 
as follows. 
\begin{enumerate}
	\item \emph{Construction of $\Phi^{(n)}$:} \ 
        Define $\Phi^{(n)}: {\cal X}^{2n}\to {\cal X}^{m}$ by
        \begin{align*}
         \Phi^{(n)}(\vck,\vcx)
        &=\varphi^{(n)}(\vck)\oplus \phi^{(n)}(\vcx)
        \\ 
        &\quad \mbox{ for } \vck, \vcx\in {\cal X}^{n}.
        \end{align*}
        Let ${C}^{m}=\Phi^{(n)}(\rvcx,\rvck)$. 
        We send ${C}^{m}$ to the public communication 
        channel. Let $\wt{X}^{m}=\phi^{(n)}(\rvcx)$
        and $\wt{K}^{m}=\varphi^{(n)}(\rvck)$. Then we have  
${C}^{m}=\wt{X}^{m}\oplus \wt{K}^{m}.$
	\item\emph{Decoding at Sink Node $\D$: \ }
	First, using the linear encoder
	$\varphi^{(n)}$,
	$\D$ encodes the key $\rvck$ received
	through private channel into
	$\widetilde{K}^{m}=$$\varphi^{(n)}({\rvck})$.
	Receiving ${C}^{m}$ from
	public communication channel, $\D$ computes
	$\widetilde{X}^{m}$ in the following way.
        Since ${C}^{m}=\wt{X}^{m}\oplus \wt{K}^{m}$, 
        the decoder $\D$ can obtain 
        $\widetilde{X}^{m}$ $=\phi^{(n)}({\rvcx})$
        by subtracting $\widetilde{K}^{m}=\varphi^{(n)}({\rvck})$ 
        from ${C}^{m}$. 
	Finally, $\D$ outputs $\hrvcx$
	by applying the decoder 
        $\psi^{(n)}$ to $\widetilde{X}^{m}$.
\end{enumerate}

\noindent
\underline{\it An Upper Bound of 
$\Delta_{{\rm max-MI}}^{(n)}(\Phi^{(n)},
         \varphi_{\cal A}^{(n)}|{p}_{KZ}^n)$:} 
We have the following upper bound of 
 $\Delta_{{\rm max-MI}}^{(n)}(\Phi^{(n)},
                  \varphi_{\cal A}^{(n)}|{p}_{KZ}^n)$. 
\begin{lemma}\label{lem:LemB} 
For the proposed construction of $\Phi^{(n)}$, 
we have 
$$
\Delta_{{\rm max-MI}}^{(n)}(\Phi^{(n)},
         \varphi_{\cal A}^{(n)}|{p}_{KZ}^n)
\leq m \log |{\cal X}|-H(\wt{K}^m| M_{\cal A}^{(n)}). 
$$
\end{lemma}
\begin{IEEEproof}
Let $\overline{X}^n $ be an arbitrary random variable 
taking values in ${\cal X}^n$. Set 
$\overline{C}^m=\Phi^{(n)}(\rvck,\overline{X}^n)$.
For the proposed construction of $\Phi^{(n)}$, we have 
$
\overline{C}^m=\wt{K}^m \oplus \phi^{(n)}(\overline{X}^n).
$
Then we have the following chain of inequalities:
\begin{align}
& I(\overline{C}^{m}; \overline{X}^n| M_{\cal A}^{(n)})
=  H(\overline{C}^{m}|M_{\cal A}^{(n)})
  -H(\overline{C}^{m};\overline{X}^n| M_{\cal A}^{(n)})
\notag\\
&= H(\overline{C}^{m} | M_{\cal A}^{(n)})
  -H(\wt{K}^m \oplus \phi^{(n)}(\overline{X}^n)
|\overline{X}^n  M_{\cal A}^{(n)})
\notag\\
&\leq m \log |{\cal X}| - H(\wt{K}^m | 
\overline{X}^n M_{\cal A}^{(n)})
\notag\\
&
= m \log |{\cal X}|-H(\wt{K}^m | M_{\cal A}^{(n)}).
\label{eqn:AsDDss}
\end{align}
Since (\ref{eqn:AsDDss}) holds for any $\overline{X}^n$, we have
the upper bound of  
$
\Delta_{{\rm max-MI}}^{(n)}(\Phi^{(n)},
         \varphi_{\cal A}^{(n)}|{p}_{KZ}^n)
$
in Lemma \ref{lem:LemB}. 
\end{IEEEproof}

\subsection{
Several Definitions
} 

In this subsection we define sets related to inner bounds 
of $\mathcal{R}_{\mathsf{Sys}}({p}_{X},{p}_{KZ})$. 
We further define functions 
related to upper bounds of 
$\Delta_{{\cal A}}^{(n)}(\Phi^{(n)},$ $R_{\cal A}|{p}_{KZ}^n)$,
which hold for $(\Phi^{(n)},\Psi^{(n)})$ proposed 
in the previous subsection. 
Let $U$ be an auxiliary random variable taking values in 
a finite set ${\cal U}$. We assume that the joint 
distribution of $(U,Z,K)$ is 
$$ 
p_{U{Z}{K}}(u,z,k)=p_{U}(u)
p_{{Z}|U}(z|u)p_{K|Z}(k|z).
$$
The above condition is equivalent to $U \markov Z \leftrightarrow K$. 
Define the set of probability distribution $p=p_{UZK}$
by
\begin{align*}
&{\cal P}(p_{KZ})
\defeq 
\{p_{UZK}: \pa {\cal U} \pa 
\leq \pa {\cal Z} \pa+1, U \markov  Z \markov K \}.
\end{align*}
Let $\mathbb{R}_{+}^2:=\{ R_{\cal A}\geq 0,R\geq 0 \}$.
Let ${\cal R}_{\rm AKW}(p_{KZ})$ be the subset of 
$\mathbb{R}_{+}^2$ such that for some $U$ with 
$p_{UZK}\in {\cal P}(p_{KZ})$,
$$
R_{\cal A}\geq I_{\empty}({Z};{U}), 
R \geq H_{\empty}({K}|{U}).
$$
The region ${\cal R}_{\rm AKW}(p_{KZ})$ 
is equal to the rate region for the one helper source coding 
problem posed and investigated by 
Ahlswede and K\"orner \cite{ahlswede:75} and Wyner \cite{wyner:75c}. 
The subscript ``AKW'' in ${\cal R}_{\rm AKW}(p_{KZ})$ is 
derived from their names. 
We can easily show that the region ${\cal R}_{\rm AKW}(p_{KZ})$ satisfies the 
following property.
\begin{property}\label{pr:pro0}  
$\quad$
\begin{itemize}
\item[a)] 
The region ${\cal R}_{\rm AKW}(p_{KZ})$ is a closed convex 
subset of $\mathbb{R}_{+}^2:=\{ R_{\cal A}\geq 0,R\geq 0 \}$.
\item[b)] The point $(0,H_{\empty}(K))$ always belongs to 
${\cal R}_{\rm AKW}(p_{KZ})$. Furthermore, for any $p_{KZ}$,
\begin{align*}
 {\cal R}_{\rm AKW}(p_{KZ}) \subseteq &
\{(R_{\cal A},R): R_{\cal A}+R \geq H_{\empty}(K)\} 
\cap \mathbb{R}_{+}^2.
\end{align*} 
\end{itemize}
\end{property}

We next explain that the region ${\cal R}_{\rm AKW}(p_{KZ})$ 
can be expressed with a family of supporting hyperplanes. 
To describe this result, we define 
a set of probability distributions on 
${\cal U}$ $\times{\cal X}$ $\times{\cal Y}$ by
\begin{align*}
{\cal P}_{\rm sh}(p_{KZ})
&\defeq 
\{p=p_{UXY}: \pa {\cal U} \pa \leq \pa {\cal Z} \pa,
  U \markov  Z \markov K \}.
\end{align*}
For $\mu\in [0,1]$, define
\begin{align*}
& R^{(\mu)}(p_{KZ})
\defeq 
\min_{p \in {\cal P}_{\rm sh}(p_{KZ})}
\left\{{\mu} I(Z;U)+ \overline{\mu}H(K|U)\right\},
\end{align*}
where $\overline{\mu}=1-\mu$. Furthermore, define
\begin{align*}
& {\cal R}_{\rm AKW, sh}(p_{KZ})
\\
&\defeq \bigcap_{\mu \in [0,1]}\{(R_{\cal A},R_{\empty}):
\ba[t]{l}
{\mu} R_{\cal A}+ \overline{\mu}R_{\empty} \geq R^{(\mu)}(p_{KZ})\}.
\ea 
\end{align*}
Then, we have the following property.
\begin{property}\label{pr:pro0z} $\quad$
\begin{itemize}
\item[a)] The bound $|{\cal U}|\leq |{\cal X}|$
is sufficient to describe 
$R^{(\mu)}($ $p_{KZ})$. 
\item[b)] For any $p_{KZ},$ we have
\beq
 {\cal R}_{\rm AKW, sh}(p_{KZ})={\cal R}_{\rm AKW}(p_{KZ}).
\label{eqn:PropEqB}
\eeq
\end{itemize}
\end{property}

Proof this property is found in Oohama \cite{oohama:19}. 
We next define a function related to an exponential 
upper bound of 
$\Delta_{\rm max-MI}^{(n)}(\Phi^{(n)},\varphi_{\A}^{(n)}$ 
$|{p}_{KZ}^n)$. 
Set
\begin{align*}
{\cal Q}(p_{K|{Z}}) \defeq& \{q=q_{U{Z}K}: 
\pa {\cal U} \pa \leq \pa {\cal {Z}} \pa,
{U} \markov {{Z}} \markov {K}, 
\\
& p_{K|{Z}}=q_{K|{Z}}\}.
\end{align*}
For $(\mu,\alpha) \in [0,1]^2$ and 
for $q=q_{U{Z}{K}}\in {\cal Q}(p_{K|{Z}})$, 
define 
\begin{align*}
& \omega_{q|p_{Z}}^{(\mu,\alpha)}({z},k|u)
 \defeq 
 \overline{\alpha} \log \frac{q_{{Z}}({z})}{p_{{Z}}({z})}
\\
&\quad + \alpha \left[
  {\mu}\log \frac{q_{{Z}|U}({z}|u)}{p_{{Z}}({z})}\right.
\left. +\overline{\mu}\log\frac{1}{q_{K|U}(k|u)}
\right],
\\
& \Omega^{(\mu,\alpha)}(q|p_{Z})
\defeq 
-\log {\rm E}_{q}
\left[\exp\left\{-\omega^{(\mu,\alpha)}_{q|p_{Z}}({Z},K|U)\right\}\right],
\\
& \Omega^{(\mu,\alpha)}(p_{KZ})
\defeq 
\min_{\scs 
   \atop{\scs 
    q \in {{\cal Q}}(p_{K|{Z}})
   }
}
\Omega^{(\mu,\alpha)}(q|p_{Z}).
\end{align*}
Furthermore, define 
\begin{align*}
& F(R_{\cal A}, R_{\empty}|p_{KZ})
\\
& \defeq \sup_{\scs (\mu,\alpha)
                                {\scs \in [0,1]^2}}
\frac{
\Omega^{(\mu,\alpha)}(p_{KZ})
-\alpha({\mu}R_{\cal A} + \overline{\mu} R_{\empty})
}
{2+\alpha \overline{\mu}}.
\end{align*}
We finally define a function serving as a lower 
bound of $F(R_{\cal A}, R_{\empty}|p_{KZ})$. 
For $\lambda \geq 0$ and for  
$p_{UXY} \in {\cal P}_{\rm sh}(p_{KZ})$, define
\begin{align*}
& \tilde{\omega}_{p}^{(\mu)}({z},k|u)
\defeq {\mu}
  \log \frac{p_{{Z}|U}({z}|u)}{p_{{Z}}({z})}
  +\log \frac{1}{p_{K|U}(K|U)},
\\
& \tilde{\Omega}^{(\mu,\lambda)}(p)
\defeq 
-\log 
{\rm E}_{p}
\left[\exp\left\{-\lambda
\tilde{\omega}_p^{(\mu)}({Z},K|U)\right\}\right],
\\
& \tilde{\Omega}^{(\mu,\lambda)}(p_{KZ})
 \defeq \min_{\scs \atop{\scs 
p \in {{\cal P}_{\rm sh}(p_{KZ})}}}
\tilde{\Omega}^{(\mu,\lambda)}(p).
\end{align*}
Furthermore, define
\begin{align*}
&{\loF}(R_{\cal A},R|p_{KZ}) 
\\
& \defeq \sup_{ \lambda \geq 0,\mu \in [0,1]} 
\frac{\tilde{\Omega}^{(\mu,\lambda)}(p_{KZ})
-\lambda({\prmtA} R_{\cal A}+{\prmtB} R)}{2+\lambda(5-{\prmtA})}.
\end{align*}
We can show that the above functions
satisfy the following property.
\begin{property}\label{pr:pro1}  
$\quad$
\begin{itemize}
\item[a)] 
The cardinality bound 
$|{\cal U}|\leq |{\cal X}|$ in ${\cal Q}(p_{K|Z})$
is sufficient to describe the quantity
$\Omega^{({\mu,\alpha})}(p_{KZ})$. 
Furthermore, the cardinality bound 
$|{\cal U}|\leq |{\cal X}|$ in ${\cal P}_{\rm sh}(p_{KZ})$
is sufficient to describe the quantity
$\tilde{\Omega}^{(\mu,\lambda)}(p_{KZ})$. 
\item[b)] For any $R_{\cal A}, R \geq0$, we have 
\begin{align*}
& &F(R_{\cal A},R_{\empty}|p_{KZ})\geq {\loF}(R_{\cal A},R_{\empty}|p_{KZ}).
\end{align*}
\item[c)] When $(R_{\cal A}+\tau, R +\tau) 
\notin {\cal R}(p_{KZ})$ for $\tau>0$, there 
exist $\lambda_0>0$ and $\mu_0 \in [0,1]$ such that
$$
{\loF}(R_{\cal A},R|p_{KZ}) > \frac{\tau}{2}
\cdot\frac{\lambda_0}{2+\lambda_0(5-\mu_0)}.
$$
\end{itemize}
\end{property}

Proofs the parts a) and b) of this property is found 
in Oohama \cite{oohama:19}. Proof of the part c) 
is given in Appendix \ref{apd:ProofProOnePartC}.

\newcommand{\ProofProOnePartC}{
\subsection{
Proof of Property \ref{pr:pro1} part c)}
\label{apd:ProofProOnePartC}

{\it Proof of Property \ref{pr:pro1} part c):} 
By simple computation we have that for any $\mu \in [0,1]$ 
and any $p \in {\cal P}_{\rm sh}(p_{KZ})$, we have the 
following:  
\begin{align}
& \lim_{\lambda \to 0}
\frac{\Omega^{(\mu,\lambda)}(p)}{\lambda}
=\left(
\frac{\rm d}{ {\rm d} \lambda } 
{\Omega}^{(\mu,\lambda)}(p)
\right)_{\lambda=0}
\notag\\
&=\mu I(U;Z) + 
\overline{\mu}
H(K|U).
\label{eqn:QSdRRR}
\end{align} 
By the hyperplane expression ${\cal R}_{\rm AKW, sh}(p_{KZ})$ of 
${\cal R}_{\rm AKW}($ $p_{KZ})$ stated Property \ref{pr:pro0z} part b)
we have that when 
$(R_{\cal A}+\tau, R+\tau) \notin {\cal R}_{\rm AKW}(p_{KZ})$, 
we have 
\begin{align} 
&{\mu}_0 R_{\cal A} + \overline{\mu_0}R < R^{(\mu_0)}(p_{KZ})-\tau
\label{eqn:XddccPP}
\end{align} 
for some $\mu_0 \in [0,1]$. We fix $p \in {\cal P}_{\rm sh}(p_{KZ})$ 
arbitrary. By (\ref{eqn:QSdRRR}), 
there exists $\lambda_0>0$ such that 
\begin{align}
& \frac{\Omega^{(\mu_0,\lambda_0)}(q)}{\lambda_0}
\geq \mu_0 I(Z;U) + \overline{\mu_0} H(K|U)-\frac{\tau}{2}
\notag\\
& \geq R^{(\mu_0)}(p_{KZ})-\frac{\tau}{2}
\MGeq{a} {\mu}_0 R_{\cal A} + \overline{\mu_0}R +\frac{\tau}{2}.
\label{eqn:ZssP}
\end{align}
Step (a) follows from (\ref{eqn:XddccPP}). 
Since (\ref{eqn:ZssP}) holds for any 
$q \in {\cal P}_{\rm sh}(p_{KZ})$, we have
\beq 
{\Omega^{(\mu_0,\lambda_0)}(p_{KZ})}
\geq 
\lambda_0\left[{\mu}_0 R_{\cal A} + \overline{\mu_0}R
+\frac{\tau}{2}\right].
\label{eqn:ZsssF}
\eeq 
Then, we have the following chain of inequalities: 
\begin{align*}
&  F(R_{\cal A},R|p_{KZ}) 
\geq \frac{
\Omega^{(\mu_0,\lambda_0)}(p_{KZ})
-\lambda_0(\mu_0 R_{\cal A} + \overline{\mu_0} R)
      }
{1+\lambda_0(5-\mu_0)}
\\
& \MG{a} \frac{1}{2} \frac{\tau \lambda_0}
{1+\lambda_0(5-\mu_0)}.
\end{align*}
Step (a) follows from (\ref{eqn:ZsssF}).
\hfill\IEEEQED
}

\subsection{Sufficient Condition for Secure Transmission} 

In this subsection we find a sufficient condition for secure 
transmission, deriving an inner bound of 
$\mathcal{R}_{\mathsf{Sys}}(\varepsilon, \delta|p_{X},p_{KZ})$. 
We first derive an explicit upper bound of 
$\Delta_{{\rm max-MI}}^{(n)}(\Phi^{(n)},$ $
 \varphi_{\cal A}^{(n)} |{p}_{KZ}^n)$. By Lemma \ref{lem:LemB}, 
It sufficies to derive an upper bound of 
$m \log |{\cal X}|-H(\wt{K}^m | M_{\cal A}^{(n)}).$
According to Santoso and Oohama \cite{santosoOh:19}, we have 
the following result.

\begin{proposition} \label{pro:ProCodingB}
$\exists \{ \varphi^{(n)} \}_{n \geq 1}$ such that
$\forall \varphi_{\A}^{(n)}$ satisfying 
$||\varphi_{\cal A}^{(n)}||\leq {\rm e}^{n R_{\cal A}}$, 
\begin{align}
& m \log |{\cal X}|-H(\wt{K}^m| M_{\cal A}^{(n)})
\leq 5nR{\ExP}^{-nF(R_{\A},R|p_{KZ})}.
\end{align}
\end{proposition}

We have two remarks on this proposition.  
\begin{remark}
Propostition \ref{pro:ProCodingB} has a close connection with the 
exponential strong converse theorem \cite{oohama:19} for the one 
helper source coding problem 
stating that the error probability of decoding outside the admissible 
rate region must tend to one exponentially as $n \to \infty$. 
In \cite{oohama:19}, an explicit form of this exponent 
function is derived. 
Proposition \ref{pro:ProCodingB} is proved 
by an application of the above result.
\end{remark}
\begin{remark}
A result similar to Proposition \ref{pro:ProCodingB} 
is obtained by Watanabe and Oohama \cite{watanabe2012privacy}. 
Let $V^m$ be the uniform random vector over ${\cal X}^m$ 
and let 
$\Delta_{\rm V}^{(n)}(p_{\hat{K}^m M_{\cal A}^{(n)}},
                                p_{M_{\cal A}^{(n)}} \times p_{V^m})$ 
$\in [0,1]$ be the normalized variational variational distance between 
$p_{\hat{K}^m M_{\cal A}^{(n)}}$ 
and $p_{M_{\cal A}^{(n)}} \times p_{V^m}$. 
Here $\hat{K}^m=f^{(n)}(K^n)$ is an image of 
the map $f^{(n)}: {\cal X}^n \to {\cal X}^m$. 
Watanabe and Oohama proved that $\forall (R_{\cal A},R)$ and 
$ \forall p_{KZ}$ satisfying 
$(R_{\cal A},R) \notin {\cal R}_{\rm AKW}(p_{KZ})$, 
$\exists \{f^{(n)} \}_{n \geq 1}$ such that
$\forall \gamma \in (0,1]$, 
$\exists n_0=n_0(\gamma) \in \mathbb{N}$, 
$\forall n\geq n_0$, 
$\forall \varphi_{\A}^{(n)}$ satisfying 
$||\varphi_{\cal A}^{(n)}||\leq {\rm e}^{n R_{\cal A}}$, 
$
\Delta_{\rm V}^{(n)}(p_{\hat{K}^m M_{\cal A}^{(n)}},
                p_{M_{\cal A}^{(n)}} \times p_{V^m}) 
\leq \gamma. 
$
To obtain this result they use the strong converse 
theorem for the one helper source coding problem 
established by Ahlswede {\it et al.} \cite{ahlswede:76}. 

\end{remark}

Combining Propositions 
\ref{pro:ProCoding}, 
\ref{pro:ProCodingB}, and 
Lemma \ref{lem:LemB}, we have the following result. 
\begin{theorem}\label{Th:mainth2}{
$\forall \gamma>0$, $\forall R_{\cal A}, \forall R>0$, and 
$\forall p_{KZ}$ with $(R_{\cal A}, R)$ 
$\in {\cal R}_{\rm AKW}^{\rm c}(p_{KZ})$, 
$\exists \{(\Phi^{(n)}, \Psi^{(n)}) \}_{n \geq 1}$
satisfying 
\begin{align*}
& \log |{\cal X}^{m}|= (m/n) \log |{\cal X}|
\in \left[R-(1/n), R\right]
\end{align*}
such that $\forall p_X$ with $R> H(X)$, 
\begin{align}
& p_{\rm e}(\phi^{(n)},\psi^{(n)}|p_{X}^n) \leq 
(n+1)^{|{\cal X}|}{\ExP}^{-nE_{\gamma}(R|p_{X})}, 
\label{eqn:mainThErrB}
\\
& \Delta_{\cal A}^{(n)}(\Phi^{(n)}, R_{\cal A}|p_{KZ}^n)
\leq 5nR{\ExP}^{-nF(R_{\A},R|p_{KZ})}. 
\label{eqn:mainThSecB}
\end{align}
}\end{theorem}

Set 
${\cal R}(p_{X},p_{KZ})
\defeq \{R \geq H(X)\} \cap {\rm cl}
[{\cal R}_{\rm AKW}^{\rm c}(p_{KZ})].
$
The functions $E_{\gamma}(R|p_{X})$ and $F(R_{\A},R|p_{KZ})$ 
take positive values if and only if $(R_{\A},R)$ belongs 
to the inner point 
of ${\cal R}(p_{X},$ $p_{KZ})$. 
By Theorem \ref{Th:mainth2}, under 
$(R_{\A},R) \in {\cal R}(p_{X},$ $p_{KZ})$,  
we have the followings: 
\begin{itemize}
\item On the reliability, 
$p_{\rm e}(\phi^{(n)},\psi^{(n)}|p_{X}^n)$ 
goes to zero exponentially as $n$ tends to infinity, and its 
exponent is lower bounded by the function $E_{\gamma}(R|p_{X})$.  
\item On the security, 
$
\Delta_{\cal A}^{(n)}(\Phi^{(n)},R_{\A}|$
$p_{KZ}^n)$ goes to zero exponentially as $n$ tends to infinity, and 
its exponent is lower bounded by the function $F(R_{\A},R|p_{KZ})$.
\item The code that attains the exponent functions 
$E_{\gamma}($ $R|p_{X})$ is the universal code that depends only on $R$ 
not on the value of the distribution $p_{X}$.
\end{itemize}

\begin{figure}[t]
\centering
\includegraphics[width=0.32\textwidth]{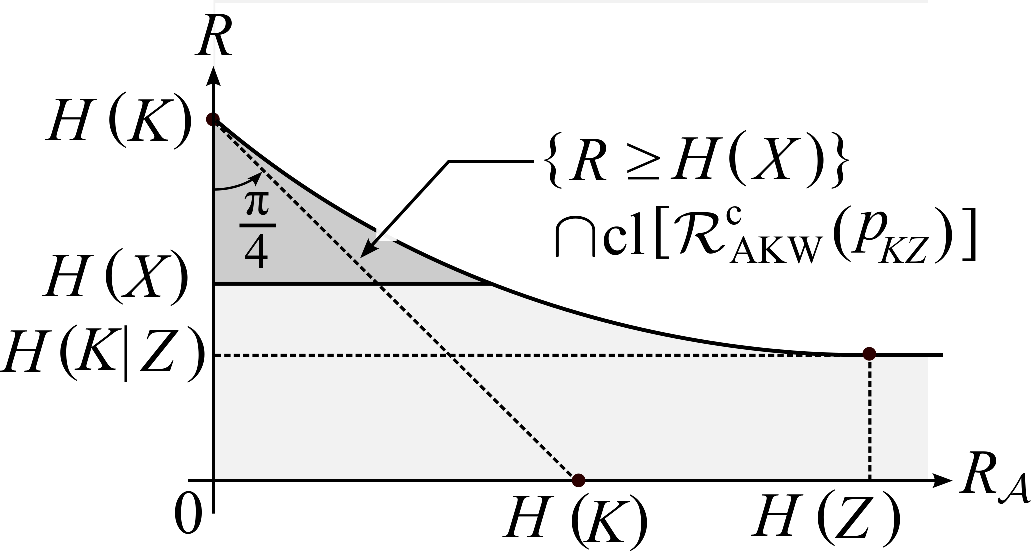}
\vspace*{-2mm}
\caption{Shape of the region 
${\cal R}(p_{X},p_{KZ})$.
}\label{fig:admissible}
\vspace*{-2mm}
\end{figure}
From Theorem \ref{Th:mainth2}, we have the following corollary. 
\begin{corollary}\label{cr:mainDirect}
\begin{align*}
&{\cal R}(p_{X},p_{KZ})
 \subseteq {\cal R}_{\rm Sys}(p_{X},p_{KZ})
\\
& \subseteq {\cal R}_{\rm Sys}(\varepsilon,\delta| p_{X},p_{KZ}).
\end{align*}
\end{corollary}
A typical shape of the region ${\cal R}(p_{X},p_{KZ})$
is shown in Fig. \ref{fig:admissible}. 

\section{
Strong Converse Theorem
}

We first derive one simple outer bound for source coding.
By the strong converse coding theorem for source coding
we have that if 
$ R < H(X)$
then 
$\forall \tau \in (0,1)$, $\forall \gamma >0$,
and $\forall \{(\phi^{(n)}$, $\psi^{(n)})\}_{n \geq 1}$,
        $\exists n_0=n_0(\tau,\gamma) \in \mathbb{N}$, 
	$\forall n\geq n_0$, we have the following:
	\begin{align*}
        & \frac{m}{n} \log |{\cal X}|\leq R + \gamma, \:
        p_{{\rm e}}(\phi^{(n)},\psi^{(n)}|p^n_{X})
        \geq 1-\tau.
	\end{align*} 
Hence we have the following theorem. 
\begin{theorem}
\label{th:SStConvTh}
For each $(\varepsilon, \delta) 
\in (0,\varepsilon_0] \times (0,1)$, we have
\begin{align}
\mathcal{R}_{\rm Sys}(\varepsilon,\delta|p_{X},p_{KZ})
\subseteq \{R\geq H(X)\}. 
\notag
\end{align} 
\end{theorem}

We next prove that for some $\varepsilon_0>0$, 
the set ${\rm cl}[\mathcal{R}_{\rm AKW}^{\rm c}($ $p_{KZ})]$ 
serves as an outer bound of 
$\mathcal{R}_{\rm Sys}(\varepsilon,\delta|p_{X},$ $p_{KZ})$ 
for $ (\varepsilon,\delta) \in (0,\varepsilon_0] \times (0,1)$.
From the definition of 
the region 
$\mathcal{R}_{\rm Sys}(\varepsilon,\delta|p_{X},p_{KZ})$ and 
Proposition {\ref{pro:ProOneA}} part b), we immediately 
obtain the following proposition.
\begin{proposition}
\label{pro:proTwo}

If $(R_{\cal A},R) \in 
\mathcal{R}_{\rm Sys}(\varepsilon,\delta|p_{X},p_{KZ})$, 
then we have that $\forall \gamma>0$, $\exists n_0(\gamma)$, 
$\forall n \geq n_0(\gamma)$, and $\forall \varphi_{\cal A}
=\{\varphi_{\cal A}^{(n)}\}_{n=1}^{\infty}$, 
\begin{align*}
& R_{\cal A} \geq \frac{1}{n}I(Z^n;\varphi_{\cal A}^{(n)}(Z^n)), 
\\
& R \leq \frac{1}{n}H(K^n|\varphi_{\cal A}^{(n)}(Z^n))+\gamma 
         +\frac{\varepsilon}{n}. 
\end{align*} 
\end{proposition}

From this proposition we have the following theorem.
\begin{theorem}
\label{th:SStConvThTwo}
For each $(\varepsilon, \delta) 
\in (0,\varepsilon_0] \times (0,1)$, we have
\begin{align*}
\mathcal{R}_{\rm Sys}(\varepsilon,\delta|p_{X},p_{KZ})
\subseteq {\rm cl}[{\cal R}_{\rm AKW}^{\rm c}(p_{KZ})].
\end{align*} 
\end{theorem}

Proof of Theorem \ref{th:SStConvThTwo} is given in 
Appendix \ref{apd:ProofStConv}. This theorem can be proved by 
Proposition \ref{pro:proTwo} and a method used in the direct 
part of the one helper helper source 
coding problem \cite{ahlswede:75}, \cite{wyner:75c}. 

\newcommand{\ProofStConv}{
\subsection{Proof of Theorem \ref{th:SStConvThTwo}}
\label{apd:ProofStConv}


\begin{IEEEproof}[Proof of Theorem \ref{th:SStConvThTwo}]
We assume that $(R_{\cal A},R) \in 
\mathcal{R}_{\rm Sys}(\varepsilon,\delta|p_{X},p_{KZ})$. 
Then by Proposition \ref{pro:proTwo}, we have 
that $\forall \gamma>0$, $\exists n_0(\gamma)$, 
$\forall n \geq n_0(\gamma)$, and $\forall \varphi_{\cal A}
=\{\varphi_{\cal A}^{(n)}\}_{n=1}^{\infty}$, 
\begin{align}
\left.
\ba{rl}
R_{\cal A}& \geq \ds \frac{1}{n}I(Z^n;\varphi_{\cal A}^{(n)}(Z^n)), 
\vspace{2mm}\\
        R & \leq \ds \frac{1}{n}H(K^n|\varphi_{\cal A}^{(n)}(Z^n))+\gamma 
                 +\frac{\varepsilon}{n}. 
\ea
\right\}
\label{eqn:Sffzzk}
\end{align} 
Here we choose any $U$ such that $p=p_{UZK} \in {\cal P}(p_{KZ})$.
Then by a method used in the proof of direct coding theorem for 
one helper source coding problem 
\cite{ahlswede:75}, \cite{wyner:75c}, 
we can show that 
$\exists \varphi_{\cal A}^{(n)}:{\cal Z}^n 
\to {\cal M}_{\cal A}$ such that 
\begin{align}
\left.
\ba{l}
\ds  \frac{1}{n}I(Z^n;\varphi_{\cal A}^{(n)}(Z^n))
\geq I(Z;U) -\nu_{1,n}, 
\vspace{2mm}\\
\ds \frac{1}{n} H(K^n|\varphi_{\cal A}^{(n)}(Z^n))
\leq H(K|U)+\nu_{2,n},  
\ea
\right\}
\label{eqn:Spdfzzk}
\end{align}
where $\left\{\nu_{i,n}\right\}_{n=1}^{\infty},i=1,2$ are 
some suitable sequences such that 
$\nu_i \to 0, n \to \infty$ for $i=1,2$.  
From (\ref{eqn:Sffzzk}) and (\ref{eqn:Spdfzzk}), 
we have 
\begin{align}
\left.
\ba{rl}
R_{\cal A} \geq & \ds I(Z;U)-\nu_{1,n},  
\vspace{2mm}\\
         R \leq & \ds H(K|U)+\nu_{2,n}
         +\gamma +\frac{\varepsilon}{n}. 
\ea
\right\}
\label{eqn:SffPP}
\end{align} 
Letting $n \to \infty$ in (\ref{eqn:SffPP}), we have 
\begin{align}
& R_{\cal A} \geq I(Z;U),\: R \leq H(K|U) + \gamma. 
 \label{eqn:ZddZZ}
\end{align}
Since $\gamma>0$ is arbitrary in (\ref{eqn:ZddZZ}), we obtain
that $\forall U$ with $p_{UZK} \in {\cal P}(p_{KZ})$, 
\begin{align}
&  R_{\cal A} \geq I(Z;U),\: R \leq H(K|U). 
 \label{eqn:Zsss}
\end{align}
Define 
$$
R_{\min}(R_{\cal A})
\defeq 
\min_{
    \scs U: p_{UZK} \in {\cal P}(p_{KZ}),
\atop{ \scs I(Z;U) \leq R_{\cal A}}}H(K|U).
$$ 
Then we have that
\begin{align*} 
&\forall U\mbox{ with }p_{UZK}\in {\cal P}(p_{KZ}),
\mbox{ we have }(\ref{eqn:Zsss})
\\
&\Longleftrightarrow
\forall U \mbox{ with }p_{UZK}\in {\cal P}(p_{KZ})
\mbox{ and }I(Z;U) \leq R_{\cal A},
\\
&\qquad \mbox{ we have }R \leq H(K|U) 
\\
& \Longleftrightarrow R \leq  R_{\min}(R_{\cal A})
  \Longleftrightarrow 
  (R_{\cal A},R) \in {\rm cl}
[{\cal R}_{\rm AKW}^{\rm c}(p_{KZ})],
\end{align*}
completing the proof.
%
\end{IEEEproof}
}

Combining Corollary \ref{cr:mainDirect}, 
Theorems \ref{th:SStConvTh}, and \ref{th:SStConvThTwo}, 
we obtain the following:
\begin{theorem}\label{th:StConvTh}
For each $(\varepsilon, \delta) 
\in (0,\varepsilon_0] \times (0,1)$, we have
\begin{align}
&     {\cal R}(p_{X},p_{KZ})=\{R\geq H(X)\}
\cap {\rm cl}[\mathcal{R}_{\mathrm {AKW}}^{\rm c}(p_{KZ})]
\notag\\
&=\mathcal{R}_{\rm Sys}(p_{X},p_{KZ}) 
=\mathcal{R}_{\rm Sys}(\varepsilon,\delta|
                      p_{X},p_{KZ}).
\notag
\end{align} 
\end{theorem}

\appendix

\ProofPrOnDecSet
\ProofPrOne
\ProofLemOne
\ProofProOneA
\ProofProOnePartC
\ProofStConv

\bibliographystyle{IEEEtran}

\bibliography{Isit2022.bib}
\end{document}

%% file: Bagus-Isit2022-abstract.tex

\begin{abstract}
	We are interested in investigating the security of source encryption
	with a symmetric key under side-channel attacks.
	In this paper, we propose a general framework of source encryption
	with a symmetric key under the side-channel attacks,
	which applies to \emph{any} source encryption
	with a symmetric key and \emph{any} kind of side-channel attacks
    targeting the secret key.
    We also propose a new security criterion for
	strong secrecy under side-channel attacks,
	which 
    is a natural extension of mutual information, i.e.,
	\emph{the maximum conditional mutual information between the plaintext
		and the ciphertext given the adversarial key leakage,
		where the maximum is taken over all possible plaintext distribution}.
	Under this new criterion, we successfully formulate the rate region,
	which serves as both necessary and sufficient conditions to have
	secure transmission even under side-channel attacks.
	Furthermore, we also prove another theoretical result on 
	our new security criterion, which might be interesting in its own right:
    in the case of the discrete memoryless source, 
    no perfect secrecy under side-channel attacks in the
    standard security criterion, i.e., the ordinary mutual information,
    is achievable  without achieving perfect
	secrecy in this new security criterion,
	although our new security criterion is more strict than
	the standard security criterion.
\end{abstract}

%% file: Bagus-Isit2022-introduction.tex

\section{Introduction \label{sec:introduction}}
As more cryptographic devices are deployed in
open physical spaces,
a new security challenge has arisen
in the form of attackers which launch
\emph{side-channel attacks}, where
an attacker does not only collect the encrypted
data sent to the public communication channel,
but also the
physical information related to the private data
which are leaked by the devices
in the form such as power consumption,
electromagnetic radiation, running time, etc.
Therefore, we consider that 
\emph{designing an encryption scheme 
that is guaranteed to be secure even under side-channel attacks} 
is very important.

In this paper, we propose a general framework
for analyzing \emph{any} source encryption
with a symmetric key under \emph{any} kind of
side-channel attacks from which 
the adversary 
obtains some leaked information on the secret key.
Although  Santoso and Oohama
investigated a similar problem in
\cite{santosoOh:19},
their work is  limited only to a single \emph{specific}
encryption scheme, i.e., one-time-pad encryption. 
In contrast, our framework here covers
\emph{any} encryption scheme.
Then, we propose a new security criterion for secrecy
which is defined as \emph{the maximum of all conditional mutual information
	between the ciphertext and plaintext given
the adversarial key leakage}.
The maximum is taken over
\emph{all probability distributions of plaintexts}.
Note that 
a security criterion is basically
a metric for representing the total amount of 
leakage of private information, i.e., the loss of secrecy.
Thus,
our new security criterion is 
more strict than the standard security criterion, i.e.,
ordinary mutual information.
Nevertheless, we show  that we can construct
a concrete encryption scheme with reliable decoding 
and secrecy under  side-channel attacks 
based on the new security criterion.
We also prove that the perfect secrecy in the new security criterion
is the necessary condition for the perfect secrecy in the standard security criterion
in the case of the discrete memoryless source.
In \cite{CsisarNarayan04}, Csisz\'{a}r and Narayan introduced
another  security criterion 
in the form of summation of mutual information and some other terms.
However, we believe that our security criterion is much more natural as an extension of mutual information compared to theirs.

The most important result in this paper is 
that we prove the \emph{strong converse}, 
i.e., the necessary condition under the new security criterion
to have an encryption scheme with both
reliable decoding and secrecy under side-channel attacks on the secret key.
At the heart of the strong converse is 
the lower bound of the amount of private information  
obtained by an adversary in any encryption scheme under side-channel attacks.
The key for deriving the lower bound 
is our main lemma which can be seen 
as an extension of Birkhoff-von Neumann 
theorem \cite{iwamoto:11}.
In a nutshell, our main lemma states that 
in \emph{any} symmetric key encryption scheme
where the plaintexts and the secret keys are 
independent, regardless of the information obtained  
via side-channel attacks on the secret key,
the adversary can see 
the encryption process as a stochastic matrix.
Based on this, we prove further that 
as long as the plaintexts are uniformly distributed, 
the adversary will see 
the ciphertexts as uniformly random,
regardless of the information on the secret key
it gets from side-channel attacks. 